\documentclass{article}
\usepackage{amssymb}
\usepackage{amsmath}

\begin{document}

\noindent \textbf{Is there a \textquotedblleft Charge - Magnet
Paradox\textquotedblright \bigskip \medskip }

\noindent Tomislav Ivezi\'{c}$\bigskip $

\noindent \textit{Ru%
\mbox
 {\it{d}\hspace{-.15em}\rule[1.25ex]{.2em}{.04ex}\hspace{-.05em}}er Bo\v{s}%
kovi\'{c} Institute, P.O.B. 180, 10002 Zagreb, Croatia}

\noindent \textit{ivezic@irb.hr}\bigskip \bigskip

In this paper it is shown that in the approach to special relativity in
which an independent physical reality is attributed to the four-dimensional
(4D) geometric quantities, the invariant special relativity (ISR), there is
not recently presented paradox that in a static electric field a magnetic
dipole moment (MDM) is subject to a torque in some frames and not in others.
Hence, in the ISR, there is no need either for the change of the Lorentz
force, but as a 4D geometric quantity, or for the introduction of some
\textquotedblleft hidden\textquotedblright\ 3D quantities. Furthermore, in
the ISR, contrary to all previous approaches, an electrically neutral
current-loop in its rest frame possesses not only a MDM $m$, but also an
electric dipole moment (EDM) $p$ and a stationary permanent magnet possesses
not only an intrinsic magnetization $M$ but also an intrinsic electric
polarization $P$. Hence, in a static electric field, both, a current-loop
and a permanent magnet experience the Lorentz force $K_{L}$ and the torque $%
N $ (bivector), i.e., the \textquotedblleft space-space\textquotedblright\
part $N_{s}$ and the \textquotedblleft time-space\textquotedblright\ part $%
N_{t}$ of $N$, in all relatively moving inertial frames. The quantities $m$,
$p$, $M$, $P$, $K_{L}$, $N$, $N_{s}$, $N_{t}$ are the 4D geometric
quantities.\bigskip

\medskip
\bigskip

\noindent \textbf{1. Introduction}\bigskip

In a recent article [1] it is argued that in the presence of the
magnetization $\mathbf{M}$ and the electric polarization $\mathbf{P}$ the
usual expression for the Lorentz force with the 3-vectors fails to accord
with the principle of relativity, because it leads to an apparent paradox
involving a magnetic dipole moment (MDM) $\mathbf{m}$ in the presence of an
electric field $\mathbf{E}$; in a static electric field a MDM $\mathbf{m}$
is subject to a torque $\mathbf{T}$ in some frames and not in others. In
this notation all 3-vectors are designated in boldface type. In [1] it is
argued that the Lorentz force should be replaced by the Einstein-Laub law,
which predicts no torque $\mathbf{T}$ in all frames. In the following, we
shall partly rely on the results and the explanations from [2]. In Section
9.1 in [2], Mansuripur's paper [1], the highlight of it [3] and some of the
critics of it [4] are considered. Mansuripur's response to the critics is
given in [5]. In [1, 3-5], \emph{all quantities} $\mathbf{E}$, $\mathbf{B}$,
$\mathbf{P}$, $\mathbf{T}$, \emph{etc. are the three-dimensional (3D)
vectors and it is considered that their transformations} (they will be
called the usual transformations (UT)) \emph{are the relativistically
correct Lorentz transformations }(LT)\emph{\ }(boosts)\emph{.} Here, in the
whole paper, under the name LT we shall only consider boosts.

Here, in Section 2, the UT of the 3-vectors $\mathbf{E}$ and $\mathbf{B}$,
Eq. (\ref{be}), and their derivation, Eq. (\ref{ei}), are objected.
According to these UT $\mathbf{E}^{\prime }$ for one inertial observer is
\textquotedblleft seen\textquotedblright\ as slightly changed $\mathbf{E}$
and an \emph{induced} $\mathbf{B}$ for a relatively moving inertial
observer. The similar UT hold for other 3-vectors, $\mathbf{P}$ and $\mathbf{%
M}$, the EDM $\mathbf{p}$ and MDM $\mathbf{m}$, see Eq. (\ref{plp}), for the
torque 3-vector $\mathbf{T}$ and another 3D vector $\mathbf{R}$, Eq. (\ref%
{te}), etc. In all papers [1, 3-5] the induced EDM $\mathbf{p}$, Eq. (\ref%
{plp}), for the moving MDM $\mathbf{m}$, leads to the interaction with the
electric field $\mathbf{E}$ and to the offending torque $\mathbf{T}$, i.e.,
to the violation of the principle of relativity and to the above mentioned
paradox. In Section 2, it is also shown that the derivation of the UT as in (%
\ref{ei}) is not possible if instead of Einstein's synchronization a
nonstandard \textquotedblleft radio,\textquotedblright\ \textquotedblleft
r\textquotedblright\ synchronization, is used, see Eqs. (\ref{are}) and (\ref%
{fe}). Also, the \textquotedblleft r\textquotedblright\ synchronization is
briefly described.

In this paper (and in Section 9.2 in [2]) it is shown that in the recently
developed geometric approach to special relativity (SR), i.e., in the
invariant SR (ISR), in which an independent physical reality is attributed
to the 4D geometric quantities and not, as usual, to the 3D quantities,
\emph{the principle of relativity is naturally satisfied and there is no
paradox.} Hence, there is no need either for the change of the expression
for the Lorentz force, but as a 4D geometric quantity, or for the
introduction of some \textquotedblleft hidden\textquotedblright\ 3D
quantities. For a brief review of the ISR see, e.g., [2] and references
therein. In some papers, e.g., [6-9], the ISR is called the
\textquotedblleft True transformations relativity.\textquotedblright\

In the ISR, which is perfectly suited to the symmetry of the 4D spacetime,
we deal either with the abstract, coordinate-free, 4D geometric quantities,
e.g., vectors (4-vectors in the usual notation) $E(x)$, $B(x)$, .. ($x$ is
the position vector), or with the 4D coordinate-based geometric quantities
(CBGQs) comprising both components and \emph{a basis}, e.g., $E=E^{\nu
}\gamma _{\nu }$. The CBGQs are invariant under the passive LT, see, e.g.,
Eq. (\ref{ecr}). Such a 4D geometric quantity represents \emph{the same
physical quantity} for relatively moving inertial observers. It is not the
case in all usual approaches that deal with the 3D quantities and their UT
or, as in the usual covariant approaches, e.g., [10], with the components of
tensors, which are implicitly taken in the standard basis, see below.

In Section 3, the primary quantity for the electromagnetic field, the
bivector $F$ is introduced and its decomposition into the vectors $E$ and $B$
is presented in Eqs. (\ref{feb}) and (\ref{fm}). The vectors $E$ and $B$
\emph{are then derived from} $F$ \emph{and} $v$, the velocity vector of the
observers who measure $E$ and $B$ fields, Eqs. (\ref{eba}) and (\ref{ebv}).
Similar equations are presented for the generalized
magnetization-polarization bivector $\mathcal{M}$ and its \textquotedblleft
time-space\textquotedblright\ and \textquotedblleft
space-space\textquotedblright\ parts, the polarization vector $P$ and the
magnetization vector $M$, respectively, Eqs. (\ref{mp}) and (\ref{pm}), for
the dipole moment bivector $D$\ and its \textquotedblleft
time-space\textquotedblright\ and \textquotedblleft
space-space\textquotedblright\ parts, the EDM vector $p$ and MDM vector $m$,
respectively, Eqs. (\ref{dm}) and (\ref{dp}), for the 4D angular momentum
bivector $J$ and its \textquotedblleft time-space\textquotedblright\ and
\textquotedblleft space-space\textquotedblright\ parts, the angular
momentum\ vectors $J_{t}$ and $J_{s}$, respectively, Eqs. (\ref{jot}) - (\ref%
{jst}), for the torque bivector $N$ and its \textquotedblleft
time-space\textquotedblright\ and \textquotedblleft
space-space\textquotedblright\ parts, the torque\ vectors $N_{t}$ and $N_{s}$%
, respectively, Eqs. (\ref{en}) - (\ref{nst}). The same equations as for $J$
and $J_{s}$,$\ J_{t}$, hold also for the spin bivector $\mathcal{S}$ as the
\emph{intrinsic} angular momentum and its usual \textquotedblleft
space-space\textquotedblright\ intrinsic angular momentum vector $S$ and its
\textquotedblleft time-space\textquotedblright\ intrinsic angular momentum
vector $Z$, but, the velocity vector $v$ of observers that appears in Eqs. (%
\ref{jot}) - (\ref{jst}) is replaced by $u$, the velocity vector of the
particle, see Section 8 in [2] and references therein. Furthermore, in
Section 3, the LT of the electric field vector are given by Eq. (\ref{el})
and their derivation is given by Eq. (\ref{et}). In the 4D spacetime, in
contrast to the UT of the 3D quantities, \emph{the mathematically correct LT
transform, e.g., the electric field vector again to the electric field
vector; there is no mixing with the magnetic field vector. The same LT hold
for all other vectors, }$x$, $P$, $M$, $p$, $m$,\ $J_{t}$, $J_{s}$, $N_{t}$,
$N_{s}$, $S$, $Z$, etc.

In this paper, the treatment of the interaction between a static electric
field and a permanent magnet will be as in Section 9.2 in [2], i.e., very
similar to the treatments of Jackson's paradox [11]\ and the Trouton-Noble
paradox [12, 13]. For simplicity, mainly the standard basis \{$\gamma _{\mu
};\ 0,1,2,3$\} of orthonormal vectors, with $\gamma _{0}$ in the forward
light cone, will be used. The $\gamma _{k}$ ($k=1,2,3$) are spacelike
vectors. As already stated, we use the term vector for the usual 4-vector,
but as a 4D geometric quantity. The $\{\gamma _{\mu }\}$ basis corresponds
to the specific system of coordinates with Einstein's synchronization [14]
of distant clocks and Cartesian space coordinates $x^{i}$. Using the 4D
geometric quantities it is shown, e.g., in [6] and [2], that an electrically
neutral current-loop (superconducting) \emph{in its rest frame} possesses
not only a MDM $m$, but also an EDM $p$. In the second paper in [6], the
incorrect quadrupole field of the \emph{stationary} current loop from the
published version is replaced by the dipole field. Similarly, it is shown in
Section 8 in [2] that a \emph{stationary} permanent magnet possesses not
only an intrinsic magnetization $M$ but also an intrinsic electric
polarization $P$. That result was derived using the generalized
Uhlenbeck-Goudsmit hypothesis from [15] according to which the connection
between the dipole moment bivector $D$ and the spin bivector $\mathcal{S}$
is given as $D=g_{S}\mathcal{S}$, Eq. (\ref{ds}). From (\ref{ds}) and using
the decompositions of $D$ (\ref{dm}) and $\mathcal{S}$, the same as (\ref%
{jms}), we write the connections between $m$ and $p$ and the corresponding
intrinsic angular momentums, vectors, $S$ and $Z$ as Eq. (\ref{sz}). Hence,
\emph{in a static electric field, both, a current-loop and a permanent
magnet experience the Lorentz force }(vector) $K_{L}$ \emph{and the torque}
(bivector) $N$ \emph{in all relatively moving inertial frames and there is
no paradox.}

In Section 4, the main results are obtained. First, the most general
expression for the Lorentz force density $k_{L}$ as an abstract vector is
presented in Eq. (\ref{kla}). In the 4D spacetime, it is relativistically
correct expression and there is no need to change it. Furthermore, if $k_{L}$
is written as a CBGQ in the standard basis it becomes Eq. (\ref{n}). In $%
S^{\prime }$, the rest frame of the magnet, it is given by the relation (\ref%
{kl}). That $k_{L}$ from (\ref{n}) and also from (\ref{kl}) is, as any other
CBGQ, invariant under the passive LT, which means that it is not as in [1,
3-5] different for relatively moving inertial observers. The integrated
torque $N$ as a CBGQ is given by Eq. (\ref{nsc}) in $S^{\prime }$ and by Eq.
(\ref{m}) in $S$, the lab frame. The expression (\ref{m}) is obtained by the
use of the LT from Eq. (\ref{nsc}). It is shown in (\ref{enc}) that $N$ in $%
S^{\prime }$, Eq. (\ref{nsc}), is equal to $N$ in $S$, Eq. (\ref{m}). This
result directly proves that with the use of the 4D torque $N$ there is no
paradox. Of course, the same result is obtained dealing with the
\textquotedblleft space-space\textquotedblright\ torque $N_{s}$ and the
\textquotedblleft time-space\textquotedblright\ torque $N_{t}$, Eqs. (\ref%
{tst}), (\ref{st}) and (\ref{nts}).

In Sections 5 - 5.6.2.2 we have discussed some other differences between the
conventional formulation of SR that deals with the 3-vectors and their UT
and this formulation, the ISR, that deals with 4D geometric quantities. In
Section 5.1, it is argued that usual Maxwell's equations (with 3-vectors),
Eqs. (1-4) in [1], have to be replaced by Eqs. (\ref{I1}) and (\ref{I2}),
which are invariant under the LT. In Section 5.2, it is briefly explained
that, contrary to the assertion from [1, 5], the Lorentz force law is
compatible with momentum conservation laws if all quantities are the 4D
geometric quantities. In Section 5.3, it is explained that Eq. (\ref{lf}) is
the relativistic version of Newton's law\ and not, as argued in [1], Eq. (%
\ref{efp}). In Section 5.4, the charge densities for a moving or stationary
infinite wire with a steady current are investigated. It is explained that
the Lorentz contraction is not well-defined in the 4D spacetime and that
only the rest length is properly defined quantity. Therefore, in the 4D
spacetime, the charge density of moving charges is meaningless and only the
current density vector $j$ is a well-defined physical quantity. Moreover,
the usual conclusion that magnetism is a relativistic phenomenon is not
correct in the 4D spacetime, because it is derived using the Lorentz
contraction and the definition of charge in terms of 3D quantities (\ref{qcl}%
). The charge defined by Eq. (\ref{qcl}) is not invariant under the LT and
in the 4D spacetime that definition has to be replaced by the
relativistically correct definition (\ref{qh}). The current density vector $%
j $, as a CBGQ in the rest frame of the wire, is determined by Eq. (\ref{swj}%
). It is mentioned that the result (\ref{swj}) leads to the existence of an
external electric field outside a stationary wire with a steady current. In
Section 5.5, another paradox from [1] is examined in detail. In [1], it is
argued that a current-carrying wire in the presence of a constant, uniform
electric field experiences a Lorentz force in some frames and not in others.
This paradox is again obtained using the 3-vectors and the UT (\ref{be}),
i.e., the UT of the 3-force. In [1], it is stated that the SR is not
violated because there is \textquotedblleft an increase in the mass of the
wire\textquotedblright\ under the action of the $E$ field, what is,
according to [1], in agreement with \textquotedblleft the relativistic
version of Newton's law,\textquotedblright\ i.e., with Eq. (\ref{efp}). In
[1], such an explanation is considered to be the resolution of the paradox.
However, it is shown in that section 5.5 that again \emph{there is no
paradox if the 4D geometric quantities are used}, i.e., if the Lorentz force
density $k_{L}$ is given by Eq. (\ref{kel}). In Sections 5.6 - 5.6.2.2, the
electromagnetic field of a point charge in uniform motion is examined. That
example very nicely illustrates the essential difference between the
conventional SR and the ISR. First, we present the expressions for $F$ as an
abstract bivector (\ref{cvf}) and as a CBGQ (\ref{cf}). That $F$ in (\ref%
{cvf}) and (\ref{cf}) depends only on the velocity vector $u_{Q}$ of the
charge $Q$ and not on $v$. It is argued in that section that the 3-vectors $%
\mathbf{E}^{\prime }$ and $\mathbf{E}$, $\mathbf{H}$ that are defined by Eq.
(11) and by Eqs. (12a), (12b) in [1], respectively, have to be replaced by $%
F $ defined by (\ref{cvf}) and its representation (\ref{cf}), i.e., by the
CBGQs defined by (\ref{q1}) and (\ref{fsq}). They are all the 4D geometric
quantities that properly transform under the LT. Then, $E$ and $B$ as
abstract vectors are given by Eq. (\ref{ebe}) and as CBGQs by Eq. (\ref{ecb}%
). $E$ and $B$ from (\ref{ebe}) and from (\ref{ecb}) depend on two velocity
vectors $u_{Q}$ and $v$. This enable us to compare the usual expressions for
the 3-vectors $\mathbf{E}^{\prime }$ and $\mathbf{E}$, $\mathbf{H}$ from [1]
with $E$, $B$ fields in the case when the observers who measure fields are
at rest, $v=c\gamma _{0}^{\prime }$, in $S^{\prime }$, the rest frame of the
charge $Q$ (Section 5.6.2.1) and in the case when these observers are in $S$%
, the lab frame, $v=c\gamma _{0}$, in which the charge $Q$ is moving
(Section 5.6.2.2). In the first case (Section 5.6.2.1), as can be seen from
Eq. (\ref{bf}), $B$ in $S^{\prime }$ is $B=B^{\prime \mu }\gamma _{\mu
}^{\prime }=0$, whereas the components $E^{\prime \mu }$ of $E$ in $%
S^{\prime }$ ($E=E^{\prime \mu }\gamma _{\mu }^{\prime }$) agree with the
usual result, e.g., with the components of the 3-vector $\mathbf{E}^{\prime
} $ given by Eq. (11) in [1]. However, as seen from Eq. (\ref{bs}), the
components $E^{\mu }$ and $B^{\mu }$ in $S$, which are obtained by the LT (%
\ref{el}) from those in (\ref{bf}), significantly differ from the components
of the 3-vectors $\mathbf{E}$ and $\mathbf{H}$ given by Eqs. (12a), (12b) in
[1], which are obtained by the UT (\ref{be}). Particularly, since the
magnetic field vector transforms by the LT again to the magnetic field
vector, $B$ remains zero for all relatively moving inertial observers. In
the second case (Section 5.6.2.2), as can be seen from Eq. (\ref{seb}),
\emph{both} 4D vector fields $E$ \emph{and} $B$ are different from zero in $%
S $, the lab frame, in which the charge $Q$\ is moving. Now, the components $%
E^{\mu }$ and $B^{\mu }$ in $S$ agree with the components of the 3-vectors $%
\mathbf{E}$ and $\mathbf{H}$ given by Eqs. (12a), (12b) in [1]. However, as
seen from Eq. (\ref{sc}), the components $E^{\prime \mu }$ and $B^{\prime
\mu }$ in $S^{\prime }$, the rest frame of the charge $Q$, which are
obtained by the LT (\ref{el}) from those in (\ref{seb}), are completely
different than the components of the 3-vectors $\mathbf{E}^{\prime }$ and $%
\mathbf{H}^{\prime }$ ($\mathbf{H}^{\prime }=0$) given by Eq. (11) in [1].
Particularly, in (\ref{sc}), $B^{\prime \mu }\gamma _{\mu }^{\prime }$ is
different from zero and, of course, the same holds for all relatively moving
inertial observers. Thus, in sections 5.6.2.1 and 5.6.2.2, it is proved that
the usual expressions with the 3-vectors, Eqs. (11), (12a) and (12b) in [1],
are not equivalent to the expressions with the 4D geometric quantities, the
abstract vectors $E$ and $B$ given by Eq. (\ref{ebe}), or CBGQs which are
given by Eq. (\ref{ecb}). This means that the usual expressions for the
electric and magnetic fields as the 3-vectors are not relativistically
correct in the 4D spacetime. Observe that all four expressions for $E$ and $%
B $, (\ref{bf}), (\ref{bs}), (\ref{seb}) and (\ref{sc}) give the same $F$
from (\ref{cf}), if they are introduced into $F$ from (\ref{fm}). Remember
that $F $ from (\ref{cvf}) and (\ref{cf}) does not contain the velocity of
the observer $v$, but only the velocity vector $u_{Q}$ of the charge $Q$.
This proves that the electromagnetic field, here the bivector $F$, is the
primary quantity for the whole electromagnetism and not the electric and
magnetic fields.

In Section 6 the conclusions are briefly exposed.

In this paper, the presentation will be in the geometric algebra formalism,
see a brief summary in Section 2 in [2] and references therein. Here, for
the reader's convenience, we shall write all equations with the CBGQs in the
standard basis and only some of them with the abstract multivectors. Hence,
\emph{the knowledge of the geometric algebra is not required for the
understanding of this presentation. }But, nevertheless, we provide a very
brief summary of the geometric algebra. The geometric (Clifford) product is
written by simply juxtaposing multivectors $AB$. The geometric product of a
grade-$r$ multivector $A_{r}$ with a grade-$s$ multivector $B_{s}$
decomposes into $A_{r}B_{s}=\left\langle AB\right\rangle _{\
r+s}+\left\langle AB\right\rangle _{\ r+s-2}...+\left\langle AB\right\rangle
_{\ \left\vert r-s\right\vert }$. The inner and outer (or exterior) products
are the lowest-grade and the highest-grade terms respectively of the above
series; $A_{r}\cdot B_{s}\equiv \left\langle AB\right\rangle _{\ \left\vert
r-s\right\vert }$ and $A_{r}\wedge B_{s}\equiv \left\langle AB\right\rangle
_{\ r+s}$. For vectors $a$ and $b$ we have: $ab=a\cdot b+a\wedge b$, where $%
a\cdot b\equiv (1/2)(ab+ba)$, $a\wedge b\equiv (1/2)(ab-ba)$. Usually the
above mentioned standard basis is introduced. The generators of the
spacetime algebra (the Clifford algebra generated by Minkowski spacetime)
are taken to be four basis vectors $\left\{ \gamma _{\mu }\right\} ,\mu
=0...3,$ satisfying $\gamma _{\mu }\cdot \gamma _{\nu }=\eta _{\mu \nu
}=diag(+---).$ The basis vectors $\gamma _{\mu }$ generate by multiplication
a complete basis for the spacetime algebra: $1$, $\gamma _{\mu }$, $\gamma
_{\mu }\wedge \gamma _{\nu }$, $\gamma _{\mu }\gamma _{5}$, $\gamma _{5}$ ($%
2^{4}=16$ independent elements). $\gamma _{5}$ is the right-handed unit
pseudoscalar, $\gamma _{5}=\gamma _{0}\wedge \gamma _{1}\wedge \gamma
_{2}\wedge \gamma _{3}$. Any multivector can be expressed as a linear
combination of these 16 basis elements of the spacetime algebra.\bigskip

\noindent \textbf{2. 3-vectors }$\mathbf{E}$, $\mathbf{B}$, $\mathbf{P}$, ..
\textbf{and their UT}\textit{\bigskip }

In all conventional formulations of the relativistic electrodynamics, e.g.,
[14], [10], the UT of the 3-vectors\emph{\ }$\mathbf{E}$ and $\mathbf{B}$,
and also $\mathbf{P}$ and $\mathbf{M}$, $\mathbf{p}$ and $\mathbf{m}$, are
considered to be the LT. These UT are given by, e.g., [14], the last
equations in \S 6., II. Electrodynamical Part, or [10], Eq. (11.149) for $%
\mathbf{E}$ and $\mathbf{B}$ or Eq. (11.148) for components implicitly taken
in the standard basis and for a boost along the $x^{1}$ axis they are
\begin{eqnarray}
E_{1} &=&E_{1}^{\prime },\ E_{2}=\gamma (E_{2}^{\prime }+\beta B_{3}^{\prime
}),\ E_{3}=\gamma (E_{3}^{\prime }-\beta B_{2}^{\prime }),  \notag \\
B_{1} &=&B_{1}^{\prime },\ B_{2}=\gamma (B_{2}^{\prime }-\beta E_{3}^{\prime
}),\ B_{3}=\gamma (B_{3}^{\prime }+\beta E_{2}^{\prime }).  \label{be}
\end{eqnarray}%
The essential feature is that, e.g., \emph{the transformed} $\mathbf{B}$
\emph{is expressed by the mixture of }$\mathbf{B}^{\prime }$ \emph{and} $%
\mathbf{E}^{\prime }$, \emph{and similarly for} $\mathbf{E}$. The same holds
for the UT of $\mathbf{P}$ and $\mathbf{M}$ and also for EDM $\mathbf{p}$
and MDM $\mathbf{m}$. Thus, if a permanent magnetization $\mathbf{M}^{\prime
}$ is viewed from a moving frame it produces an electric polarization $%
\mathbf{P=\gamma U\times M}^{\prime }/c^{2}$. It is also argued that a
neutral stationary current loop with a magnetic moment $\mathbf{m}^{\prime }$
in its rest frame $S^{\prime }$, acquires an electric dipole moment
\begin{equation}
\mathbf{p}=\boldsymbol{\beta \times m}^{\prime }/c  \label{plp}
\end{equation}%
if it is moving with uniform 3-velocity $\mathbf{U}$ ($\boldsymbol{\beta =U}%
/c$ ) relative to the laboratory frame $S$. It is always assumed that \emph{%
in the rest frame of the neutral current loop the electric moment} $\mathbf{p%
}^{\prime }$ \emph{is zero,} $\mathbf{p}^{\prime }=\mathbf{0}$. For more
detail see Sections 3.1 and 3.2 in [2] and references therein. It is
visible, e.g., from Griffiths and Hnizdo (GH) [4], that the offending torque
$\mathbf{T}$ from their Eq. (5) is obtained from the term $\mathbf{p\times E}
$, where $\mathbf{p}$ is the induced EDM, their Eq. (4), which means that
\emph{the violation of the principle of relativity is a direct consequence
of the UT for the 3-vectors} $\mathbf{p}$ \emph{and} $\mathbf{m}$, i.e., $%
\mathbf{P}$ and $\mathbf{M}$. \emph{In all papers in} [1, 4] \emph{and} [5]
the same induced EDM $\mathbf{p}$ appears as a result of the UT for\emph{\ }$%
\mathbf{p}$ and $\mathbf{m}$.

In [10], the six independent components of the electromagnetic field tensor $%
F^{\alpha \beta }$ (only components in the standard basis) are identified to
be six components of the 3-vectors $\mathbf{E}$ and $\mathbf{B}$ in \emph{%
both} relatively moving inertial frames of reference,%
\begin{equation}
E_{i}^{\prime }=F^{^{\prime }i0},\ E_{i}=F^{i0};\quad B_{i}^{\prime
}=(1/2c)\varepsilon _{ikl}F^{\prime lk},\ B_{i}=(1/2c)\varepsilon
_{ikl}F^{lk}.  \label{ei}
\end{equation}%
This means that \emph{the UT of the components of} $\mathbf{E}$ \emph{and} $%
\mathbf{B}$ \emph{are derived assuming that they transform under the LT as
the components of} $F^{\alpha \beta }$ \emph{transform}, Eq. (11.148) in
[10], i.e., Eq. (\ref{be}) here. (Note that in (\ref{ei}) the components of
the 3-vectors,\ here $\mathbf{E}^{\prime }$, $\mathbf{E}$ and $\mathbf{B}%
^{\prime }$, $\mathbf{B}$ are written with lowered (generic) subscripts,
since they are not the spatial components of the 4D quantities. This refers
to the third-rank antisymmetric $\varepsilon $ tensor too. The super- and
subscripts are used only on the components of the 4D quantities.) By the
same procedure one finds the UT of the components of $\mathbf{P}$ and $%
\mathbf{M}$, of EDM $\mathbf{p}$ and MDM $\mathbf{m}$, simply replacing $%
\mathbf{E}$ and $\mathbf{B}$\ by $\mathbf{P}$ and $\mathbf{M}$, $\mathbf{p}$
and $\mathbf{m}$, and $F^{\alpha \beta }$ by the generalized
magnetization-polarization tensor $\mathcal{M}^{\alpha \beta }$, the dipole
moment tensor $D^{\alpha \beta }$ (only components in the standard basis).
Similarly, the components of the torque 3-vector $\mathbf{T}=\mathbf{r}%
\times \mathbf{F}$ and of another 3D vector $\mathbf{R}$ are identified with
the \textquotedblleft space-space\textquotedblright\ and \textquotedblleft
time-space\textquotedblright\ components respectively of the torque
four-tensor $N^{\alpha \beta }$\ in both relatively moving inertial frames
of reference, see Cross [4] and Section 9.1 in [2]. This yields the UT for
the components $T_{i}$ of $\mathbf{T}$ and $R_{i}$\ of $\mathbf{R}$, which,
as can be seen from (\ref{be}), are the same as the UT for $B_{i}$ and $%
E_{i} $, respectively.%
\begin{eqnarray}
T_{1} &=&T_{1}^{\prime },\ T_{2}=\gamma (T_{2}^{\prime }-\beta R_{3}^{\prime
}),\ T_{3}=\gamma (T_{3}^{\prime }+\beta R_{2}^{\prime }),  \notag \\
R_{1} &=&R_{1}^{\prime },\ R_{2}=\gamma (R_{2}^{\prime }+\beta T_{3}^{\prime
}),\ R_{3}=\gamma (R_{3}^{\prime }-\beta T_{2}^{\prime }).  \label{te}
\end{eqnarray}%
Here, they are written for the motion along the $x^{1}$ axis. Note that the
component in (\ref{te}) that corresponds to $T^{\prime x}$ in Cross [4] is $%
T_{2}=-\gamma \beta R_{3}^{\prime }$. \emph{The components} $T_{i}$ \emph{in
the moving frame} \emph{are expressed by the mixture of the components of }$%
\mathbf{T}^{\prime }$ \emph{and} \emph{of }$\mathbf{R}^{\prime }$ \emph{from
the rest frame. This causes that the components of} $\mathbf{T}$\emph{\ will
not vanish in the} $S$ \emph{frame even if they vanish in the} $S^{\prime }$
\emph{frame},\emph{\ i.e., that there is a \textquotedblleft charge-magnet
paradox\textquotedblright\ in all usual approaches to SR that deal with the
3-vectors or with components implicitly taken in the standard basis. }The
same holds for the components of the angular momentum $\mathbf{L}$, $\mathbf{%
L}=\mathbf{r}\times \mathbf{p}$, with $\mathbf{T}=d\mathbf{L}/dt$, and of
another 3-vector $\mathbf{K}$. It is assumed that they transform as the
\textquotedblleft space-space\textquotedblright\ and \textquotedblleft
time-space\textquotedblright\ components respectively of the usual covariant
angular momentum four-tensor $J^{\alpha \beta }$, see [16] and Section 3 in
[11]. These UT of the components of $\mathbf{L}$ are the same as the UT (\ref%
{te}) but with $L_{i}$, $K_{i}$ replacing $T_{i}$, $R_{i}$, respectively.
Observe that in [16], and also in [1, 3-5], only the \textquotedblleft
space-space\textquotedblright\ components of $J^{\alpha \beta }$ ($L_{i}$)
and $N^{\alpha \beta }$\ ($T_{i}$) are considered to be the physical angular
momentum and torque respectively, because they are associated with actual
rotation in the 3D space of the object. On the other hand, the
\textquotedblleft time-space\textquotedblright\ components of $J^{\alpha
\beta }$ ($K_{i}$) and $N^{\alpha \beta }$\ ($R_{i}$) are \emph{not}
considered to be of the same physical nature as $L_{i}$ and $T_{i}$. In all
usual treatments it is considered that $K_{i}$ and $R_{i}$ are not the
physical angular momentum and torque respectively, because they are \emph{not%
} associated with any overt rotation in the 3D space of the object, see,
particularly, GH [4] and Jackson's paper [16].

However, if one does not use Einstein's synchronization but, e.g., a
drastically different \textquotedblleft radio,\textquotedblright\
\textquotedblleft r\textquotedblright\ synchronization, then, as will be
shown below, it is not possible to make the identification of the components
of, e.g., the 3-vectors $\mathbf{E}$ and $\mathbf{B}$ with the components $%
F^{\alpha \beta }$, i.e., $L_{i}$, $K_{i}$ with the components $J^{\alpha
\beta }$, or $T_{i}$, $R_{i}$ with the components $N^{\alpha \beta }$. For
the \textquotedblleft r\textquotedblright\ synchronization, see also Section
3.1 in [2] and references therein. The \textquotedblleft
r\textquotedblright\ synchronization is commonly used in everyday life and
not Einstein's synchronization. If the observers who are at \emph{different
distances} from the studio clock set their clocks by the announcement from
the studio\ then they have synchronized their clocks with the studio clock
according to the \textquotedblleft r\textquotedblright\ synchronization.
Thus, there is an absolute simultaneity in the \textquotedblleft
r\textquotedblright\ synchronization. If, e.g., the components $F^{\alpha
\beta }$ of $F$ are transformed by the transformation matrix $R_{\;\nu
}^{\mu }$ to the $\{r_{\mu }\}$ basis, then it is obtained that
\begin{equation}
F_{r}^{10}=F^{10}-F^{12}-F^{13}.  \label{are}
\end{equation}%
The same equation holds for $J_{r}^{10}$, $N_{r}^{10}$, ... . In the
transformation matrix $R_{\;\nu }^{\mu }$, which connects the components
from the $\left\{ \gamma _{\mu }\right\} $ basis with the components from
the $\{r_{\mu }\}$ basis, the only components that are different from zero
are%
\begin{equation}
R_{\;\mu }^{\mu }=-R_{\;i}^{0}=1.  \label{er}
\end{equation}%
It is visible from (\ref{are}) that the \textquotedblleft
time-space\textquotedblright\ component in the $\{r_{\mu }\}$ basis is
expressed by the mixture of the \textquotedblleft
time-space\textquotedblright\ component and the \textquotedblleft
space-space\textquotedblright\ components from the $\{\gamma _{\mu }\}$
basis. Hence, in the $\left\{ r_{\mu }\right\} $ basis (with the
\textquotedblleft r\textquotedblright\ synchronization) it holds that $%
F_{r}^{10}=E_{1}+cB_{3}-cB_{2}$. The same identification as in $\left\{
\gamma _{\mu }\right\} $ basis, yields that, e.g., the component $E_{1r}$ in
the $\left\{ r_{\mu }\right\} $ basis is expressed as the combination of $%
E_{i}$ and $B_{i}$ components from the $\left\{ \gamma _{\mu }\right\} $
basis,%
\begin{equation}
E_{1r}=F_{r}^{10}=E_{1}+cB_{3}-cB_{2}.  \label{fe}
\end{equation}%
This means that the usual identifications, e.g., Eq. (11.137) in [10], or
those ones in Cross [4], are meaningful \emph{only} if the $\left\{ \gamma
_{\mu }\right\} $ basis is chosen, i.e., if the Minkowski metric is used.
But, \emph{every synchronization is only a convention and physics must not
depend on conventions. }

We note that in the $\left\{ r_{\mu }\right\} $ basis the usual time and
space components of the position vector $x$ cannot be separated. The
connections between the components of $x$ in both bases are given by the
relations%
\begin{equation}
x_{r}^{0}=x^{0}-x^{1}-x^{2}-x^{3},\quad x_{r}^{i}=x^{i}.  \label{ptr}
\end{equation}%
It is visible from (\ref{ptr}) that in the $\left\{ r_{\mu }\right\} $ basis
the space-time split (3+1) of the 4D spacetime is not possible. Hence,
specifically, in the 4D spacetime, the usual translation in the 3D space has
not some definite physical meaning. However, if the 4D geometric quantities
are used, i.e., \emph{in the ISR, the physics does not depend on conventions}%
, since the abstract geometric quantity, the vector $x$, can be decomposed
in both bases and it holds that%
\begin{equation}
x=x^{\mu }\gamma _{\mu }=x_{r}^{\mu }r_{\mu }.  \label{rx}
\end{equation}

Thus, as already stated in Section 1, in the 4D spacetime the physical
quantities are not correctly represented only by components, but a basis
must be included.

Furthermore, in the 4D spacetime, it is meaningless to consider that, e.g.,
the 3D $\mathbf{T}$ and $\mathbf{L}$ are physical torque and angular
momentum, respectively, whereas $\mathbf{R}$ and $\mathbf{K}$ are not of the
same physical nature. In the same way as in (\ref{fe}), the
\textquotedblleft time-space\textquotedblright\ component of $J_{r}^{\alpha
\beta }$ ($K_{1r}$) (or the same for $N_{r}^{\alpha \beta }$\ ($R_{1r}$)) in
the $\left\{ r_{\mu }\right\} $ basis will be expressed in terms of the
\textquotedblleft time-space\textquotedblright\ component ($K_{1}$) and the
\textquotedblleft space-space\textquotedblright\ components ($L_{i}$) from
the $\{\gamma _{\mu }\}$ basis. Hence, in the 4D spacetime, as mentioned
above for the usual translation in the 3D space, the usual rotation, i.e.,
an overt rotation in the 3D space has not a definite physical meaning. \emph{%
In the 4D spacetime, the correctly defined 4D angular momentum is the
bivector} $J$ \emph{given by Eq.} (\ref{jot}) \emph{and the correctly
defined 4D torque is the bivector} $N$ \emph{given by Eq.} (\ref{en}), \emph{%
which are connected by the relation} (\ref{nj}), $N=dJ/d\tau $. \emph{In the
4D spacetime, they completely describe all phenomena connected with a
rotation.}\bigskip

\noindent \textbf{3. Vectors }$E$, $B$, $P$, .. \textbf{and their LT}\textit{%
\bigskip }

Recently, in [17-22], it is proved both in the tensor formalism and in the
geometric algebra formalism, that \emph{the UT of the 3-vectors }$\mathbf{E}$
\emph{and} $\mathbf{B}$ \emph{ARE NOT the LT and also the correct LT of the
4D geometric quantities that represent the electric and magnetic fields are
derived.} In the 4D spacetime, \emph{the correct LT always transform the 4D
algebraic object representing, e.g., the electric field only to the electric
field; there is no mixing with the magnetic field. }For a review see [2]. In
[17-22], see also Section 5 in [2], the LT of the components $E^{\mu }$ (in
the $\{\gamma _{\mu }\}$ basis) of the vector $E=E^{\mu }\gamma _{\mu }$ are
given as
\begin{equation}
E^{\prime 0}=\gamma (E^{0}-\beta E^{1}),\ E^{\prime 1}=\gamma (E^{1}-\beta
E^{0}),\ E^{\prime 2,3}=E^{2,3},  \label{el}
\end{equation}%
for a boost along the $x^{1}$ axis. As mentioned above any CBGQ is unchanged
under the LT, i.e., it holds that
\begin{equation}
E=E^{\nu }\gamma _{\nu }=E^{\prime \nu }\gamma _{\nu }^{\prime }=E_{r}^{\nu
}r_{\nu }=E_{r}^{\prime \nu }r_{\nu }^{\prime },  \label{ecr}
\end{equation}%
where the primed quantities in both bases $\{\gamma _{\mu }\}$ and $\{r_{\mu
}\}$ are the Lorentz transforms of the unprimed ones; for the LT in the $%
\{r_{\mu }\}$ basis see [7]. The same LT hold for any other vector, e.g., $x$%
, $B$, $P$, $M$, EDM $p$ and MDM $m$, the torque vectors $N_{s}$ and $N_{t}$%
, see below, etc. The equation (\ref{ecr}) shows that in the 4D spacetime
the vector $E$ \emph{is the same 4D quantity} for all relatively moving
inertial frames of reference and for all systems of coordinates chosen in
them. This is an essential difference relative to all usual approaches with
the 3-vectors and their UT; the 3-vectors $\mathbf{B}^{\prime }$ and $%
\mathbf{B}$ that are connected by the UT (\ref{be}) are completely different
quantities in the 4D spacetime. Note that the same holds for the usual
covariant approach, e.g., from [10], that deals with components implicitly
taken in the standard basis, i.e., $F^{\prime \alpha \beta }\neq F^{\alpha
\beta }$; they are different quantities in the 4D spacetime. The components
do not contain the whole information about some physical quantity; a basis
must be included.

A short derivation of the LT (\ref{el}) is presented in [22] and it will be
briefly repeated here. It is proved in [12] that in the 4D spacetime the
primary physical quantity for the whole electromagnetism is the
electromagnetic field (bivector) $F$. There, an axiomatic geometric
formulation of electromagnetism is developed in which there is only one
axiom, the field equation for $F$%
\begin{equation}
\partial \cdot F+\partial \wedge F=j/\varepsilon _{0}c,  \label{df}
\end{equation}%
i.e., with the CBGQs in the $\{\gamma _{\mu }\}$ basis, that equation becomes

\begin{equation}
\partial _{\alpha }F^{\alpha \beta }\gamma _{\beta }-\partial _{\alpha }\
^{\ast }F^{\alpha \beta }\gamma _{5}\gamma _{\beta }=(1/\varepsilon
_{0}c)j^{\beta }\gamma _{\beta },  \label{c1}
\end{equation}%
where $\varepsilon ^{\alpha \beta \gamma \delta }$ is the totally
skew-symmetric Levi-Civita pseudotensor and $\gamma _{5}$, as already
stated, is the right-handed unit pseudoscalar $\gamma _{5}=\gamma _{0}\wedge
\gamma _{1}\wedge \gamma _{2}\wedge \gamma _{3}$ and $^{\ast }F^{\alpha
\beta }=(1/2)\varepsilon ^{\alpha \beta \gamma \delta }F_{\gamma \delta }$
is the usual dual tensor. The usual covariant form of Eq. (\ref{c1}), i.e.,
\emph{only the basis components in the} $\left\{ \gamma _{\mu }\right\} $
\emph{basis,} are two equations, the equation with sources $\partial
_{\alpha }F^{a\beta }=j^{\beta }/\varepsilon _{0}c$, and that one without
sources $\partial _{\alpha }\ ^{\ast }F^{\alpha \beta }=0$. It can be seen
from [12] that the bivector field $F$ yields the complete description of the
electromagnetic field. For the given sources the expression for $F$ can be
found from Eqs. (7) and (8) in [12] and there is no need to introduce either
the field vectors or the potentials. However, the field vectors can be
introduced using a mathematical theorem that any second rank antisymmetric
tensor can be decomposed into two space-like vectors and a unit time-like
vector (the velocity vector/c). Hence, $F$ can be decomposed as
\begin{equation}
F=(1/c)E\wedge v+(IB)\cdot v,  \label{feb}
\end{equation}%
where the unit pseudoscalar $I$ is defined algebraically without introducing
any reference frame. If $I$ is represented in the $\left\{ \gamma _{\mu
}\right\} $ basis it becomes $I=\gamma _{0}\wedge \gamma _{1}\wedge \gamma
_{2}\wedge \gamma _{3}=\gamma _{5}$. If that equation for $F$ is written
with the CBGQs in the $\{\gamma _{\mu }\}$ basis it becomes%
\begin{equation}
F=(1/2)F^{\mu \nu }\gamma _{\mu }\wedge \gamma _{\nu },\ F^{\mu \nu
}=(1/c)(E^{\mu }v^{\nu }-E^{\nu }v^{\mu })+\varepsilon ^{\mu \nu \alpha
\beta }v_{\alpha }B_{\beta },  \label{fm}
\end{equation}%
where $\gamma _{\mu }\wedge \gamma _{\nu }$ is the bivector basis. Observe
that bivector $F$ \emph{is the same 4D quantity} for relatively moving
inertial observers and for all bases chosen by them, i.e., the relation like
(\ref{ecr}) holds for $F$ as well, $F=(1/2)F^{\mu \nu }\gamma _{\mu }\wedge
\gamma _{\nu }=(1/2)F^{\prime \mu \nu }\gamma _{\mu }^{\prime }\wedge \gamma
_{\nu }^{\prime }=...$ . The vectors $E$ and $B$ \emph{are defined in terms
of} $F$ \emph{and} $v$, the velocity vector of a family of observers who
measure $E$ and $B$ fields in the following way
\begin{equation}
E=(1/c)F\cdot v,\quad B=-(1/c^{2})I(F\wedge v).  \label{eba}
\end{equation}%
We write them as the CBGQs in the $\{\gamma _{\mu }\}$ basis%
\begin{equation}
E=E^{\mu }\gamma _{\mu }=(1/c)F^{\mu \nu }v_{\nu }\gamma _{\mu },\quad
B=B^{\mu }\gamma _{\mu }=(1/2c^{2})\varepsilon ^{\mu \nu \alpha \beta
}F_{\nu \alpha }v_{\beta }\gamma _{\mu }.  \label{ebv}
\end{equation}%
Since $F$ is antisymmetric it holds that $E^{\mu }v_{\mu }=B^{\mu }v_{\mu
}=0 $, only three components of $E$ and $B$ in any basis are independent.
However, it does not mean that three spatial components of $E$, or $B$, are
necessarily independent components. Namely $E$ and $B$ depend not only on $F$
but on $v$ as well. The form of $v$ in a given inertial frame will determine
which three components are independent. These definitions are mathematically
correct definitions, which are first given (only in the component form) by
Minkowski in Section 11.6 in [23] and reinvented and generalized in terms of
4D geometric quantities in [17-22]. In [24], a mathematical argument is
presented according to which the electric and magnetic fields have to be
represented by the 4D geometric quantities, i.e., as in Minkowski's Section
11.6 and in [17-22]. Namely, it is explained in [24] that the number of
variables on which a vector field depends, i.e., the dimension of its domain
is essential for the number of components of that vector field. Hence, the
usual time-dependent $\mathbf{E(r,}t\mathbf{)}$, $\mathbf{B(r,}t\mathbf{)}$
cannot be the 3-vectors, since they are defined on the spacetime. Therefore,
we use the term \textquotedblleft vector\textquotedblright\ for the
correctly defined geometric quantity, which is defined on the spacetime,
e.g., $E(x)$, $B(x)$, $P(x)$, $M(x)$, etc. An incorrect expression, the
3-vector or the 3D vector, will still remain for the usual $\mathbf{E(r,}t%
\mathbf{)}$, $\mathbf{B(r,}t\mathbf{)}$ from the conventional formulations
of the electromagnetism, e.g., [14], [10], [25-30], ... . However, it has to
be noted that in the 4D spacetime we \emph{always} have to deal with
correctly defined vectors $E(x)$, $B(x)$, $P(x)$, $M(x)$, etc. even in the
usual static case, i.e., if the usual 3D fields $\mathbf{E(\mathbf{r})}$, $%
\mathbf{B(r)}$ do not explicitly depend on the time $t$. The reason is that
\emph{in the 4D spacetime there is no static case, i.e., there is no
electrostatic and magnetostatic.} The LT mix the time and space components,
which means that the LT cannot transform the spatial coordinates from one
frame only to spatial coordinates in a relatively moving inertial frame of
reference. What is static case for one inertial observer is not more static
case for relatively moving inertial observer, but a time dependent case.
Furthermore, as explained above in Section 2, if an observer uses the
\textquotedblleft r\textquotedblright\ synchronization and not the standard
Einstein's synchronization, then as seen from (\ref{ptr}) the space and time
are not separated and the usual 3D vector $\mathbf{\mathbf{r}}$ is
meaningless. Hence, if the principle of relativity has to be satisfied and
the physics must be the same for all inertial observers and for $\{\gamma
_{\mu }\}$, $\{r_{\mu }\}$, $\{\gamma _{\mu }^{\prime }\}$, etc. bases which
they use, then the properly defined quantity is the position vector $x$, $%
x=x^{\nu }\gamma _{\nu }=x^{\prime \nu }\gamma _{\nu }^{\prime }=x_{r}^{\nu
}r_{\nu }=x_{r}^{\prime \nu }r_{\nu }^{\prime }$, and not $\mathbf{\mathbf{r}%
}$ and $t$. Consequently, in the 4D spacetime, e.g., the electric field is
properly defined as the vector $E(x)$ for which the relation (\ref{ecr})
holds.

The frame of \textquotedblleft fiducial\textquotedblright\ observers for
which $v=c\gamma _{0}$, with the $\{\gamma _{\mu }\}$ basis in it, will be
called the $\gamma _{0}$-frame. This is not any kind of a preferred frame,
because any inertial frame can be chosen to be the $\gamma _{0}$-frame. In
the $\gamma _{0}$-frame $E^{0}=B^{0}=0$ and $E^{i}=F^{i0}$, $%
B^{i}=(1/2c)\varepsilon ^{0ijk}F_{kj}$; the same components as in, e.g., Eq.
(11.137) in [10]. However, in any other inertial frame, the
\textquotedblleft fiducial\textquotedblright\ observers are moving, and $%
v=v^{0}\gamma _{0}=c\gamma _{0}=v^{\prime \mu }\gamma _{\mu }^{\prime }$.
For the \textquotedblleft fiducial\textquotedblright\ observers, $v^{\mu
}=c\gamma _{0}^{\mu }$ and $E^{\mu }=F^{\mu \nu }\gamma _{0,\nu }$. It is
proved by Minkowski in Section 11.6 in [23], and reinvented in [17-22], see
also Section 5 in [2], that in the mathematically correct procedure for the
derivation of the LT of $E$ \emph{both} $F$ \emph{and} \emph{the velocity
vector }$v$ have to be transformed by the LT, e.g., as shown in [22], for
the LT from the $\gamma _{0}$-frame%
\begin{equation}
E=E^{\mu }\gamma _{\mu }=[(1/c)F^{i0}v_{0}]\gamma _{i}=E^{\prime \mu }\gamma
_{\mu }^{\prime }=[(1/c)F^{\prime \mu \nu }v_{\nu }^{\prime }]\gamma _{\mu
}^{\prime }.  \label{et}
\end{equation}%
Hence, \emph{the components} $E^{\mu }$ \emph{transform by the LT again to
the components} $E^{\prime \mu }$ \emph{of the same electric field vector},
i.e., the above quoted LT (\ref{el}) of the components $E^{\prime \mu }$ are
obtained. \emph{The main point is that the transformed components} $%
E^{\prime \mu }$ \emph{are not determined only by} $F^{\prime \mu \nu }$,%
\emph{\ as in all usual approaches}, e.g., Eqs. (11.147) and (11.148) in
[10], \emph{but also by} $v^{\prime \mu }$. In the third paper in [24]
Oziewicz, from the mathematical point of view, nicely explains the results
obtained in my papers [17-22]. (The references in the quoted part refer to
the mentioned Oziewicz's paper.) He states: \textquotedblleft Minkowski [1],
and then Ivezi\'{c} [7-10], observed correctly that if a Lorentz
transformation is an isomorphism of a vector space, then the entire algebra
of tensor fields must be Lorentz-covariant. ..... An active Lorentz
transformation must act on all tensor fields, including an observer's
time-like vector field.\textquotedblright\ This means that if the Lorentz
transformation is applied to $E$ from (\ref{eba}), $E=(1/c)F\cdot v$, then,
from the mathematical point of view, it is necessary that the Lorentz
transformation acts not only on $F$ \emph{but on} $v$, i.e.,
\textquotedblleft an observer's time-like vector field\textquotedblright ,
\emph{as well}.

It can be easily checked, see Section 6 in [2], that the UT, Eq. (11.148) in
[10], i.e., Eq. (\ref{be}) here, will be obtained if \emph{only} the
components $F^{\mu \nu }$ are transformed \emph{but not} the components $%
v^{\mu }$. Hence, from the above mentioned Oziewicz's mathematical argument
the UT cannot be - the LT.

Similarly, the bivector $\mathcal{M}$ can be decomposed as%
\begin{equation}
\mathcal{M}=P\wedge u/c+(MI)\cdot u/c^{2},  \label{M1}
\end{equation}%
or, as a CBGQ, it is written as%
\begin{equation}
\mathcal{M}=(1/2)\mathcal{M}^{\mu \nu }\gamma _{\mu }\wedge \gamma _{\nu },\
\mathcal{M}^{\mu \nu }=(1/c)(P^{\mu }u^{\nu }-P^{\nu }u^{\mu
})+(1/c^{2})\varepsilon ^{\mu \nu \alpha \beta }M_{\alpha }u_{\beta }.
\label{mp}
\end{equation}%
The vectors $P(x)$ and $M(x)$ are determined by $\mathcal{M(}x\mathcal{)}$
and the unit time-like vector $u/c$, where $u$ is identified with bulk
velocity vector of the medium in spacetime
\begin{equation}
P=(1/c)\mathcal{M}^{\mu \nu }u_{\nu }\gamma _{\mu },\quad M=(1/2)\varepsilon
^{\mu \nu \alpha \beta }\mathcal{M}_{\alpha \nu }u_{\beta }\gamma _{\mu },
\label{pm}
\end{equation}%
with $P^{\mu }u_{\mu }=M^{\mu }u_{\mu }=0$. It is visible from (\ref{pm})
that $P$ and $M$ depend not only on $\mathcal{M}$ \emph{but} \emph{on} $u$
\emph{as well}, see also Section 4 in [2]. In the same way, the bivector $D$
as the primary physical quantity for the dipole moments can be written as%
\begin{equation}
D=(1/2)D^{\mu \nu }\gamma _{\mu }\wedge \gamma _{\nu },\ D^{\mu \nu
}=(1/c)(p^{\mu }u^{\nu }-p^{\nu }u^{\mu })+(1/c^{2})\varepsilon ^{\mu \nu
\alpha \beta }m_{\alpha }u_{\beta },  \label{dm}
\end{equation}%
see also [22] and Section 4 in [15]. Then, one finds that $p$ and $m$ \emph{%
are determined by the bivector} $D$ \emph{and the velocity vector of the
particle} $u$ as%
\begin{equation}
p=(1/c)D^{\mu \nu }u_{\nu }\gamma _{\mu },\quad m=(1/2)\varepsilon ^{\mu \nu
\alpha \beta }D_{\alpha \nu }u_{\beta }\gamma _{\mu },  \label{dp}
\end{equation}%
with $p^{\mu }u_{\mu }=m^{\mu }u_{\mu }=0$. In the particle's rest frame
(the $S^{\prime }$ frame) and the $\{\gamma _{\mu }^{\prime }\}$ basis, $%
u=c\gamma _{0}^{\prime }$, which yields that $p^{\prime 0}=m^{\prime 0}=0$, $%
p^{\prime i}=D^{\prime i0}$, $m^{\prime i}=(c/2)\varepsilon
^{0ijk}D_{jk}^{\prime }$. Therefore $p$ and $m$ can be called the
\textquotedblleft time-space\textquotedblright\ part and the
\textquotedblleft space-space\textquotedblright\ part, respectively, of $D$.
The quotation marks are written because the relation, e.g., $p^{\prime
i}=D^{\prime i0}$, holds \emph{only} in the $\{\gamma _{\mu }\}$ basis but
not in other bases, e.g., in the $\left\{ r_{\mu }\right\} $ basis.

In the 4D spacetime, the primary physical quantity for the 4D angular
momentums is the bivector $J$,%
\begin{equation}
J=(1/2)J^{\mu \nu }\gamma _{\mu }\wedge \gamma _{\nu },\quad J^{\mu \nu
}=x^{\mu }p^{\nu }-x^{\nu }p^{\mu }.  \label{jot}
\end{equation}%
It can be decomposed into the \textquotedblleft
space-space\textquotedblright\ and the \textquotedblleft
time-space\textquotedblright\ angular momentum\ vectors $J_{s}$ and $J_{t}$
respectively and the velocity vector $v$ of a family of observers who
measures $J_{s}$ and $J_{t}$. The components $J^{\mu \nu }$ are given as

\begin{equation}
J^{\mu \nu }=(1/c)[(J_{t}^{\mu }v^{\nu }-J_{t}^{\nu }v^{\mu })+\varepsilon
^{\mu \nu \alpha \beta }J_{s,\alpha }v_{\beta }].  \label{jms}
\end{equation}%
Then, the vectors $J_{s}$ and $J_{t}$ \emph{are derived from} $J$ \emph{and
the velocity vector} $v$ as

\begin{equation}
J_{t}=(1/c)J^{\mu \nu }v_{\nu }\gamma _{\mu },\quad J_{s}=(1/2c)\varepsilon
^{\mu \nu \alpha \beta }J_{\alpha \nu }v_{\beta }\gamma _{\mu },  \label{jst}
\end{equation}%
with $J_{s}^{\mu }v_{\mu }=J_{t}^{\mu }v_{\mu }=0$. $J_{s}$ and $J_{t}$
depend not only on $J$ \emph{but also on} $v$. In the $\gamma _{0}$-frame $%
J_{s}^{0}=J_{t}^{0}=0$ and only the spatial components remain $%
J_{s}^{i}=(1/2)\varepsilon ^{0ijk}J_{jk}$, $J_{t}^{i}=J^{i0}$. $J_{s}^{i}$
and $J_{t}^{i}$ correspond to the components of the 3-vectors $\mathbf{L}$
and $\mathbf{K}$ that are introduced, e.g., in [16]. However, as already
stated above, in [16], as in all usual treatments including [1, 4, 10,
25-30], it is considered that only $\mathbf{L}$ is a physical quantity whose
components transform according to the UT, e.g., Eq. (11) in [16], i.e., the
same as (\ref{te}) but with $L_{i}$, $K_{i}$ replacing $T_{i}$, $R_{i}$,
respectively.

Furthermore, see [11-13] and Sections 9.1 and 9.2 in [2], the torque
bivector $N$ as a CBGQ is given as
\begin{equation}
N=(1/2)N^{\mu \nu }\gamma _{\mu }\wedge \gamma _{\nu },\quad N^{\mu \nu
}=x^{\mu }K_{L}^{\nu }-x^{\nu }K_{L}^{\mu },  \label{en}
\end{equation}%
where $K_{L}$ is the Lorentz force vector. The decomposition of $N$ into the
\textquotedblleft space-space\textquotedblright\ and the \textquotedblleft
time-space\textquotedblright\ vectors $N_{s}$ and $N_{t}$ respectively is
given as

\begin{equation}
N^{\mu \nu }=(1/c)[(N_{t}^{\mu }v^{\nu }-N_{t}^{\nu }v^{\mu })+\varepsilon
^{\mu \nu \alpha \beta }N_{s,\alpha }v_{\beta }].  \label{nd}
\end{equation}%
The \textquotedblleft time-space\textquotedblright\ torque $N_{t}$ and the
\textquotedblleft space-space\textquotedblright\ torque $N_{s}$%
\begin{equation}
N_{t}=(1/c)N^{\mu \nu }v_{\nu }\gamma _{\mu },\quad N_{s}=(1/2c)\varepsilon
^{\mu \nu \alpha \beta }N_{\alpha \nu }v_{\beta }\gamma _{\mu }  \label{nst}
\end{equation}%
are determined by $N$ \emph{and the velocity vector} $v$ of a family of
observers who measures $N_{s}$ and $N_{t}$. It holds that $N_{s}^{\mu
}v_{\mu }=N_{t}^{\mu }v_{\mu }=0$. In the $\gamma _{0}$-frame, $v^{\mu
}=(c,0,0,0)$, $N_{s}^{0}=N_{t}^{0}=0$ and \emph{only the spatial components}
$N_{s}^{i}$ and $N_{t}^{i}$ remain, $N_{s}^{i}=(1/2)\varepsilon
^{0ijk}N_{jk} $,$\ N_{t}^{i}=N^{i0}$. \emph{Both vectors} $N_{s}$ \emph{and}
$N_{t}$ \emph{are in the same measure physical 4D torques, which, only if
taken together,} \emph{are equivalent to the 4D torque} $N$. $N$ is
connected with the angular momentum bivector $J$ as
\begin{equation}
N=dJ/d\tau ,  \label{nj}
\end{equation}%
where $\tau $ is the proper time.

In the following we shall also need an important relation, the generalized
Uhlenbeck-Goudsmit hypothesis, which is explained in detail in [15]. In the
same way as $J$ is the primary physical quantity for the 4D angular
momentums the spin bivector $\mathcal{S}$ (four-tensor $S^{ab}$ in [15]) is
the primary quantity \emph{with} \emph{definite physical reality} for the
\emph{intrinsic} angular momentums. It can be decomposed into the usual
\textquotedblleft space-space\textquotedblright\ intrinsic angular momentum
vector $S$, the \textquotedblleft time-space\textquotedblright\ intrinsic
angular momentum vector $Z$ and the unit time-like vector $u/c$, where $u$
is the velocity vector of the particle. The relations are the same as (\ref%
{jms}) and (\ref{jst}), but $J$, $J_{s}$, $J_{t}$ and $v$ are replaced by $%
\mathcal{S}$, $S$, $Z$ and $u$, respectively. Then, in [15], the usual
connection between the 3-vectors $\mathbf{m}$ and $\mathbf{S}$, $\mathbf{m}%
=\gamma _{S}\mathbf{S}$, is formulated as the generalized Uhlenbeck-Goudsmit
hypothesis; the dipole moment bivector $D$ is proportional to the spin
bivector $\mathcal{S}$%
\begin{equation}
D=g_{S}\mathcal{S}.  \label{ds}
\end{equation}%
Furthermore, in [15], using the decompositions of $D$ (\ref{dm}) and $%
\mathcal{S}$, the same as (\ref{jms}), we have formulated the connections
between the dipole moments, vectors, $m$ and $p$ and the corresponding
intrinsic angular momentums, vectors, $S$ and $Z$, respectively, as%
\begin{equation}
m=cg_{S}S,\quad p=g_{S}Z.  \label{sz}
\end{equation}%
Hence, in [15], a fundamentally new result is obtained only from a
relativistically correct treatment of physical quantities $D$ and $\mathcal{S%
}$, i.e., that any fundamental particle has not only the usual spin vector $%
S $ and the corresponding intrinsic MDM $m$, but also another spin vector $Z$
and the corresponding intrinsic EDM $p$, whose magnitude is $(1/c)$ of that
for $m$. In the particle's rest frame and the $\{\gamma _{\mu }^{\prime }\}$
basis, $u=c\gamma _{0}^{\prime }$ and $p^{\prime 0}=m^{\prime 0}=0$, $%
p^{\prime i}=g_{S}Z^{\prime i}$, $m^{\prime i}=cg_{S}S^{\prime i}$. Hence,
comparing this last relation with $\mathbf{m}=\gamma _{S}\mathbf{S}$, it is
visible that $g_{S}=\gamma _{S}/c$. As shown in Section 8 in [2] these
results yield that in the same way as the MDMs determine the magnetization $%
M $ of a stationary permanent magnet the EDMs determine its polarization $P$%
, which induces an electric field outside a permanent magnet, moving or
\emph{stationary}.

This discussion explicitly shows that from the ISR viewpoint the derivation
of the transformations of the 3-vectors $\mathbf{E}$ and $\mathbf{B}$ from
[10] is not mathematically correct, i.e., Eqs. (11.148), (11.149) in [10]
are not the LT but the UT. The same holds for the UT of the 3-vectors $%
\mathbf{P}$ and $\mathbf{M}$, $\mathbf{p}$ and $\mathbf{m}$, $\mathbf{R}$
and $\mathbf{T}$, $\mathbf{K}$ and $\mathbf{L}$. Hence, from the ISR
viewpoint, all \textquotedblleft resolutions\textquotedblright\ of
Mansuripur's paradox from [1] and [3-5] are not relativistically correct,
because they are based on the use of the 3D quantities and their UT, see
also Section 9.1 in [2]. In addition, it is worth mentioning that \emph{all
treatments from} [1, 3-5] \emph{are meaningless} \emph{if only the Einstein
synchronization is replaced by the \textquotedblleft r\textquotedblright }\
\emph{synchronization}. This conclusion simply follows already from the
above mentioned equations for $F_{r}^{10}$ (\ref{are}) and the expression
for $E_{1r}$ (\ref{fe}).\bigskip

\noindent \textbf{4. With the 4D torques there is no \textquotedblleft
Charge - Magnet Paradox\textquotedblright \bigskip }

We consider the system from [1], but, without loss of generality, the
electric charge will be substituted by a uniform electric field. The common
rest frame of the source of the electric field (a point charge $q$ in [1],
i.e., $Q$ in this paper) and of the permanent magnet will be denoted as $%
S^{\prime }$, whereas the lab frame, in which the $S^{\prime }$ frame moves
with uniform velocity $V=V\gamma _{1}$ along the common $x^{1}$, $x^{\prime
1}$ axes, will be denoted as $S$. Hence, in $S^{\prime }$, only the
component $F^{\prime 10}$ ($E^{\prime 1}=F^{\prime 10}$) of $F^{\prime \mu
\nu }$ is different from zero. From the point of view of the ISR it would be
more appropriate to exclusively deal with the primary quantity $F$, i.e., $%
F^{\prime \mu \nu }$, as in the treatment of the Trouton-Noble paradox in
[12]. However, for the reader's convenience and for an easier comparison
with the usual treatments from [1] and [3-5] we shall explicitly work with
quantities that are derived by a correct mathematical procedure from $F$ and
$v$ according to (\ref{ebv}), i.e., with the vectors $E$ and $B$. Observe,
as shown in Section 5.6 below, that $E$ and $B$ are different for different
choices of the velocity vector of the observer $v$, but $F$ is the same for
all these choices of $v$. $F$ is independent of $v$ and that fact shows that
$F$ is the primary quantity for the electromagnetism and not the electric
and magnetic fields. This is completely different than in all conventional
formulations of the electromagnetism, e.g., [1], [3-5], [10], [16], [25 -
30], in which the 3-vectors $\mathbf{E}$ and $\mathbf{B}$, i.e., their
components implicitly taken in the standard basis, are considered to be the
primary quantities, whereas the components $F^{\mu \nu }$ are determined in
terms of $E_{x,y,z}$ and $B_{x,y,z}$, i.e., as already stated, they are
identified to be six components of the 3-vectors $\mathbf{E}$ and $\mathbf{B}
$ in \emph{all }relatively moving inertial frames of reference, according to
(\ref{ei}). Obviously, the same consideration holds for the bivector $%
\mathcal{M}$ as the primary quantity and the vectors $P$ and $M$ which are
derived from $\mathcal{M}$ and $u$ according to (\ref{pm}), then for the
torque $N$ as the primary quantity for the 4D torques and the vectors $N_{s}$
and $N_{t}$ which are derived from $N$ and $v$ according to (\ref{nst}), as
shown in [11], etc.

The usual expressions for the Lorentz force density $k_{L}$ as an abstract
vector and as a CBGQ are%
\begin{equation}
k_{L}=(1/c)F\cdot j,\quad k_{L}=(1/c)F^{\mu \nu }j_{\nu }\gamma _{\mu },
\label{LoF}
\end{equation}%
where the total current density vector $j$ is $j=j^{(C)}+j^{(\mathcal{M})}$;
$j^{(C)}$ is the conduction current density of the \emph{free} charges and $%
j^{(\mathcal{M})}$ is the magnetization-polarization current density of the
\emph{bound} charges, $j^{(\mathcal{M})}=-c\partial \mathcal{M}=-c\partial
\cdot \mathcal{M}$ ($\partial \wedge \mathcal{M}=0$, since $j^{(\mathcal{M}%
)} $ is a vector). If written as a CBGQ $j^{(\mathcal{M})}$ is

\begin{equation}
j^{(\mathcal{M})}=-c\partial _{\mu }\mathcal{M}^{\mu \nu }\gamma _{\nu }.
\label{jm}
\end{equation}%
In the considered case it is taken that $j^{(C)}=0$. Using the
decompositions of $F$ (\ref{feb}) and of $M$ (\ref{M1}), $k_{L}$, as an
abstract vector, becomes%
\begin{equation}
k_{L}=(1/c^{2})(E\wedge v)\cdot \lbrack -(\partial \cdot P)u+(u\cdot
\partial )P+(1/c)[u\wedge (\partial \wedge M)]I],  \label{kla}
\end{equation}%
where it is taken in the decomposition of $F$ (\ref{feb}) that in the
considered case $B=0$. In contrast to all previous treatments with the UT,
according to the LT $B$ is always $=0$ and therefore there is no reason for
the appearance of the paradox. It is visible that the expression for $k_{L}$
contains two velocity vectors, $v$ - the velocity vector of the observers
who measure $E$ and $B$ fields (from (\ref{feb}), i.e., (\ref{fm})) and $u$
- the velocity vector of the permanent magnet, i.e., of the electric current
loop (from (\ref{mp})). This $k_{L}$ is relativistically correct expression
and it does not need any change. If $k_{L}$ is written as a CBGQ in the
standard basis it becomes%
\begin{eqnarray}
k_{L} &=&(1/c^{2})\{(\partial _{\mu }P^{\mu })[(E^{\nu }u_{\nu })v^{\rho
}-(v^{\nu }u_{\nu })E^{\rho }]-(u^{\mu }\partial _{\mu })P^{\nu }[E_{\nu
}v^{\rho }-v_{\nu }E^{\rho }]  \notag \\
&&+(1/c)\varepsilon ^{\mu \nu \alpha \beta }u_{\mu }(\partial _{\alpha
}M_{\beta })[E_{\nu }v^{\rho }-v_{\nu }E^{\rho }]\}\gamma _{\rho
}=k_{L}^{\rho }\gamma _{\rho }.  \label{n}
\end{eqnarray}%
As any other CBGQ $k_{L}$ from (\ref{n}) is invariant under the LT; it is
the same 4D quantity for all relatively moving inertial observers. Here, we
write $k_{L}$ in the $S^{\prime }$ frame, i. e., for the case that $%
u=v=c\gamma _{0}^{\prime }$ and accordingly that $E^{\prime 0}=P^{\prime
0}=M^{\prime 0}=0$. Then, $k_{L}$ as a CBGQ in the $\{\gamma _{\mu }^{\prime
}\}$ basis is
\begin{equation}
k_{L}=(-E^{\prime k}\partial _{0}^{\prime }P_{k}^{\prime }+(1/c)\varepsilon
^{0jkl}E_{j}^{\prime }\partial _{k}^{\prime }M_{l}^{\prime })\gamma
_{0}^{\prime }-E^{\prime i}(\partial _{k}^{\prime }P^{\prime k})\gamma
_{i}^{\prime }.  \label{kl}
\end{equation}%
In the usual approaches with the 3-vectors and their UT, e.g., in [1] and in
GH [4], the Lorentz 3-force density is zero in the $S^{\prime }$ frame;
there is no $\gamma _{0}^{\prime }$ term and there is no $\mathbf{P}$. The
components $k_{L}^{\prime \mu }$ correspond to the time and spatial
components of $f^{\alpha }$ from Cross [4], i.e., to $f^{0}=(1/c)\mathbf{%
E\cdot }(\partial \mathbf{P}/\partial t+\mathbf{\nabla }\times \mathbf{M)}$
and $f^{i}=-E^{i}(\mathbf{\nabla P})$. But, the components $k_{L}^{\prime
\mu }$ are multiplied by the unit basis vectors $\gamma _{\mu }^{\prime }$
in order to form the CBGQ $k_{L}^{\prime \mu }\gamma _{\mu }^{\prime }$,
i.e., a representation in the standard basis of a vector $k_{L}$, whereas
the 3-vector, e.g., $\mathbf{E}$ is constructed from the components $%
E_{x,y,z}$ and \emph{the unit 3-vectors} $\mathbf{i}$, $\mathbf{j}$, $%
\mathbf{k}$. \emph{In the 4D spacetime there is no room for the 3-vectors;
they cannot correctly transform under the LT.} It is not correct to write
the components of some 4D CBGQ in terms of the 3-vectors like in Cross [4],
Vanzella [4], Barnett [4], etc. as in almost all textbooks that treat SR,
e.g., [10, 25 - 30].

For the sake of brevity we shall explicitly write the results for $N$, $%
N_{s} $ and $N_{t}$, for others see Section 9.2 in [2]. The torque density $%
n $ is $n=(1/2)n^{\mu \nu }\gamma _{\mu }\wedge \gamma _{\nu }$,$\ n^{\mu
\nu }=x^{\mu }k_{L}^{\nu }-x^{\nu }k_{L}^{\mu }$, where $k_{L}^{\mu }$\ is
given by Eq. (\ref{n}). $n$, as a CBGQ in the standard basis is given by Eq.
(69) in [2]. In the $S^{\prime }$ frame it is given by Eq. (70) in [2] and
it is $\neq 0$, whereas in the approaches with the 3-vectors $n$ is zero in
the $S^{\prime }$ frame.

In the common rest frame $S^{\prime }$, in which $v^{\prime \mu }=(c,0,0,0)$%
, the integrated torque $N$ as a CBGQ is given as

\begin{equation}
N=-(1/c)E^{\prime 1}m^{\prime 2}(\gamma _{0}^{\prime }\wedge \gamma
_{3}^{\prime })-E^{\prime 1}p^{\prime 3}(\gamma _{1}^{\prime }\wedge \gamma
_{3}^{\prime }),  \label{nsc}
\end{equation}%
where $m$ is the MDM vector and $p$ is the EDM vector. In the considered
case, $E=E^{\prime 1}\gamma _{1}^{\prime }$, $m=m^{\prime 2}\gamma
_{2}^{\prime }$ and $p=p^{\prime 3}\gamma _{3}^{\prime }$. The quantities in
(\ref{nsc}) are all properly defined in the 4D spacetime and they properly
transform under the LT. Hence, in $S$, the lab frame, the torque $N$ can be
obtained by the LT from $S^{\prime }$ and it is

\begin{equation}
N=(-E^{1}m^{2}/c+\beta E^{1}p^{3})(\gamma _{0}\wedge \gamma _{3})+(\beta
E^{1}m^{2}/c-E^{1}p^{3})(\gamma _{1}\wedge \gamma _{3}).  \label{m}
\end{equation}%
The LT of the components of the electric field vector $E$, Eq. (\ref{el}),
are used to derive that $E^{\prime 1}=(1/\gamma )E^{1}$. It can be seen from
Eqs. (\ref{nsc}) and (\ref{m}) that \emph{the 4D torque} $N$ \emph{is the
same 4D geometric quantity for all relatively moving inertial observers},
i.e., it holds that

\begin{equation}
N=(1/2)N^{^{\prime }\mu \nu }\gamma _{\mu }^{\prime }\wedge \gamma _{\nu
}^{\prime }=(1/2)N^{\mu \nu }\gamma _{\mu }\wedge \gamma _{\nu }  \label{enc}
\end{equation}%
\emph{and there is no paradox.} In the same way as in (\ref{ecr}) for $E$
the bivector $N$ will be the same 4D quantity for all bases, e.g., the $%
\{r_{\mu }\}$ basis, and not only for the standard basis.

Let us determine $N_{s}$ and $N_{t}$, which are \emph{both} equally well
physical, as 4D CBGQs in the common rest frame $S^{\prime }$. They are%
\begin{equation}
N_{s}=N_{s}^{\prime \mu }\gamma _{\mu }^{\prime }=(1/c)E^{\prime 1}p^{\prime
3}v^{\prime 0}\gamma _{2}^{\prime },\quad N_{t}=N_{t}^{\prime \mu }\gamma
_{\mu }^{\prime }=-(1/c^{2})E^{\prime 1}m^{\prime 2}v^{\prime 0}\gamma
_{3}^{\prime }.  \label{tst}
\end{equation}%
$N_{t}$ in (\ref{tst}) corresponds to the expression $\mathbf{R}=\mathbf{m}%
\times \mathbf{E}=-mE\widehat{\mathbf{y}}$ in Cross [4] that describes the
interaction of the magnetic moment with the electric field in the rest frame
$S^{\prime }$. Remember that in Cross [4] the rest frame is with unprimed
quantities and the motion is along the $x^{3}$ axis. Moreover, Cross deals
with components $N^{\mu \nu }$ and $J^{\mu \nu }$ implicitly taken in the
standard basis and not with the CBGQs. Contrary to his assertion, the
transformations of his $\mathbf{R}$ and $\mathbf{T}$, i.e., of the
corresponding \emph{components} $N^{0i}$ and $N^{ij}$ respectively are not
the LT but the UT given by Eq. (\ref{te}). Thus his $\mathbf{T}$ and $%
\mathbf{R}$ are not relativistically correct and they are completely
different than $N_{s}$ and $N_{t}$. It is visible from (\ref{nsc}) that $%
N_{t}$ in (\ref{tst}) comes from the \textquotedblleft
time-space\textquotedblright\ component $N^{\prime 03}$ and in this
geometric approach it exists even if $p^{\prime \mu }$ would be zero. $N_{s}$
in (\ref{tst}) does not appear in any previous paper since it emerges from
the existence of the EDM $p$ for a stationary permanent magnet. It describes
the interaction of the EDM $p$ of the \emph{stationary} permanent magnet
with the electric field $E$ in the rest frame $S^{\prime }$. It comes from
the \textquotedblleft space-space\textquotedblright\ component $N^{\prime
13} $ in (\ref{nsc}). In the usual formulation with the 3-vectors it would
correspond to the usual 3D torque $\mathbf{T}=\mathbf{p}\times \mathbf{E}$,
but, in contrast to all previous formulations, this torque is in the rest
frame $S^{\prime }$. Then, we determine $N_{s}$ and $N_{t}$ in $S$, e.g.,
using the LT\emph{\ }of \emph{all} quantities which determine $N_{s}$ and $%
N_{t}$ in (\ref{tst}),%
\begin{equation}
N_{s}=N_{s}^{\mu }\gamma _{\mu }=(1/\gamma )E^{1}p^{3}\gamma _{2},\quad
N_{t}=-(1/c\gamma )E^{1}m^{2}\gamma _{3}.  \label{st}
\end{equation}%
Observe that $N_{s}$ ($N_{t}$) transforms under the LT as every vector
transforms, i.e., as in (\ref{el}), which means that \emph{components} $%
N_{s}^{\prime \mu }$ \emph{of }$N_{s}$ \emph{transform to the components} $%
N_{s}^{\mu }$ of the same torque $N_{s}$ in the $S$ frame; \emph{there is no
mixing with the components of }$N_{t}$. These LT of the components of $N_{s}$
($N_{t}$) are obtained in the same way as the LT of $B$ ($E$) are obtained,
i.e., that \emph{both} $N$ \emph{and} $v$ from the definitions of $N_{s}$ ($%
N_{t}$) (\ref{nst}) are transformed by the LT. This is in a sharp contrast
to the UT of the 3D torque $\mathbf{T}$, Eq. (\ref{te}), in which the
transformed components $T_{i}$ are expressed by the mixture of components $%
T_{k}^{\prime }$ of the 3-vector $\mathbf{T}^{\prime }$ and of components $%
R_{k}^{\prime }$ of another 3-vector $\mathbf{R}^{\prime }$ from the rest
frame. These UT of $T_{i}$ (and $R_{i}$) can be obtained in such a way that
\emph{only} $N$ from the definitions of $N_{s}$ ($N_{t}$) (\ref{nst}) is
transformed by the LT, but \emph{not} the velocity of the observer $v$.

It is worth noting that the CBGQs $N_{s}^{\prime \mu }\gamma _{\mu }^{\prime
}$ and $N_{s}^{\mu }\gamma _{\mu }$ are \emph{the same quantity} $N_{s}$ in $%
S^{\prime }$ and $S$ frames, and the same for $N_{t}$,
\begin{equation}
N_{s}=N_{s}^{\prime \mu }\gamma _{\mu }^{\prime }=N_{s}^{\mu }\gamma _{\mu
},\quad N_{t}=N_{t}^{\prime \mu }\gamma _{\mu }^{\prime }=N_{t}^{\mu }\gamma
_{\mu }.  \label{nts}
\end{equation}%
The relation (\ref{nts}) holds in the same measure for all bases and not
only for the standard basis, as in (\ref{ecr}). This again shows, as in
[11-13], that \emph{in the approach with the 4D torques} $N_{s}$ \emph{and} $%
N_{t}$ \emph{the principle of relativity is naturally satisfied and there is
no paradox}. Observe that $N_{s}$ is always determined by the interaction of
the EDM $p$ and $E$, whereas $N_{t}$ is determined by the interaction of $m$
and $E$. In this geometric approach there is no need either for the
\textquotedblleft hidden\textquotedblright\ 3D mechanical angular momentum
or for the \textquotedblleft hidden\textquotedblright\ 3D torque.

Let us examine what would be if it is taken that, as in the usual treatments
[1, 3-5], a permanent magnet possesses only a MDM $m$ and not an EDM $p$.
Note, that in our approach there is $p\neq 0$ and the assumption that $p=0$
is only taken for some comparison with the usual approaches. However, it is
worth mentioning that even in this case $p=0$ we deal with correctly defined
vectors in the 4D spacetime and with their LT and not with the 3-vectors and
their UT. Then, for $p=0$, as can be seen from (\ref{nsc}), the
\textquotedblleft space-space\textquotedblright\ component $N^{\prime 13}$
is zero, but the \textquotedblleft time-space\textquotedblright\ component $%
N^{\prime 03}$ is different from zero. As already stated, in the
conventional treatments only the \textquotedblleft
space-space\textquotedblright\ components $N^{\prime ij}$ are considered to
be physical, i.e., that they are three components of the 3D torque $\mathbf{T%
}^{\prime }$, which is connected with the usual 3D rotation. But, in this
geometric approach, as explained at the end of Section 2, the usual rotation
in the 3D space has not a definite physical meaning. In the 4D spacetime,
only the whole $N$ given by (\ref{nsc}) does have a definite physical
reality. In $S$, the torque $N$ is given by Eq. (\ref{m}) and for $p=0$ both
the \textquotedblleft space-space\textquotedblright\ component $N^{13}$ and
the \textquotedblleft time-space\textquotedblright\ component $N^{03}$ are
different from zero. They are both determined by the interaction of the
magnetic moment $m$ with the electric field $E$. As can be easily seen, Eq. (%
\ref{enc}) holds in this case too and \emph{there is no paradox}.

It is visible from (\ref{tst}) and (\ref{st}) that for $p=0$ the
\textquotedblleft space-space\textquotedblright\ part of $N$, the torque
vector $N_{s}$ is always zero, $N_{s}=0$, but $N_{t}$ is different from
zero, $N_{t}\neq 0$, and it is always the same 4D geometric quantity, which
means that again \emph{there is no paradox}. Remember that only if $N_{s}$
and $N_{t}$ are taken together then they are equivalent to the primary
physical quantity for the 4D torques, to the bivector $N$. It is, as
explained above, different from zero and it is the same 4D quantity for all
relatively moving inertial frames of reference. Also, as in the case with $%
p\neq 0$, there is no need for the change of the expression for the Lorentz
force, but as a 4D geometric quantity, or for the introduction of some
\textquotedblleft hidden\textquotedblright\ 3D quantities.\bigskip

\noindent \textbf{5. Another differences in the treatments with 4D geometric}

\textbf{quantities and with the 3D quantities\bigskip }

The preceding consideration clearly shows that, as in the case with
Jackson's paradox [11]\ and the Trouton-Noble paradox [12, 13], Mansuripur's
paradox with the torque appears because of the use of the 3D quantities and
their UT. But, it is visible from the relations for $N$ (\ref{nsc}), (\ref{m}%
), (\ref{enc}) and those for $N_{s}$ and $N_{t}$ (\ref{tst}), (\ref{st}), (%
\ref{nts}) that \emph{there is no paradox if an independent physical reality
is attributed to the 4D geometric quantities and if their LT are used.}
According to that in the ISR there is no need to introduce some
\textquotedblleft hidden\textquotedblright\ quantities. These
\textquotedblleft hidden\textquotedblright\ quantities are introduced in
different ways in almost all papers in [4]. In the 4D spacetime,\emph{\ }%
they are without well-defined physical meaning. Simply, they are an artifact
of the use of the 3D quantities and their UT.

In the ISR, it is proved that there is a true agreement, independent of the
chosen inertial reference frame and of the chosen system of coordinates in
it, with different experiments, e.g., the motional electromotive force in
[18], the Faraday disk in [19], the Trouton-Noble experiment in [12, 13] and
also in [8, 9], the well-known experiments that test SR: the
\textquotedblleft muon\textquotedblright\ experiment, the Michelson-Morley -
type experiments, the Kennedy-Thorndike - type experiments and the
Ives-Stilwell - type experiments. \emph{This true agreement with experiments
directly proves the physical reality of the 4D geometric quantities.} It is
also shown in the mentioned papers that the agreement between the
experiments and Einstein's formulation of SR [14] \emph{is not a true
agreement} since it depends on the chosen synchronization. Remember, as
already stated several times, the conventional SR deals with the
synchronously defined spatial length , i.e., the Lorentz contraction, see
also Appendix in [2], then with the conventional dilation of time and also
with the UT of the components of the 3-vectors $\mathbf{E}$ and $\mathbf{B}$%
, Particularly, this is explicitly shown in [9] in which \emph{both}
Einstein's synchronization and the \textquotedblleft r\textquotedblright\
synchronization are used throughout the paper. As can be seen from [6-9],
contrary to the generally accepted opinion in the conventional SR, the\emph{%
\ relativity of simultaneity, the Lorentz contraction and the time dilation
are not well-defined in the 4D spacetime. }They are not the intrinsic
relativistic effects since they depend on the chosen synchronization.
However, as already stated, \emph{every synchronization is only a convention
and physics must not depend on conventions.}

In the following we shall examine several other ambiguities in the
conventional treatments of SR and present how they are removed in this
approach with the 4D geometric quantities.\emph{\bigskip }

\noindent \textit{5.1 Electromagnetic field equations for moving
media\bigskip }

In all usual approaches, e.g., [10], [25-30], including [1, 3-5], it is
supposed that Maxwell's equations with the 3-vectors, both in the vacuum and
in a moving medium, are covariant under the LT. However, for the vacuum, it
is proved in [19] that it is not true, because the transformations of the
3-vectors $\mathbf{E}$ and $\mathbf{B}$ are not the LT but the
relativistically incorrect UT and the Lorentz invariant field equations with
$E$ and $B$ are presented, Eqs. (39) and (40) in that paper (Eqs. (27-29) in
[2]). Moreover, in [12], an axiomatic geometric formulation of
electromagnetism in vacuum is developed which has \emph{only one axiom}, the
field equation for $F$, Eq. (\ref{df}) here. If it is written with the CBGQs
in the standard basis it becomes Eq. (\ref{c1}). Its generalization to a
magnetized and polarized moving medium with $\mathcal{M(}x\mathcal{)}$ is
presented in [31]. It is%
\begin{equation}
\partial (\varepsilon _{0}F+\mathcal{M})=j^{(C)}/c;\quad \partial \cdot
(\varepsilon _{0}F+\mathcal{M})=j^{(C)}/c,\ \partial \wedge F=0.  \label{F4}
\end{equation}%
If written with the CBGQs in the standard basis that equation becomes

\begin{equation}
\partial _{\alpha }(\varepsilon _{0}F^{\alpha \beta }+\mathcal{M}^{\alpha
\beta })\gamma _{\beta }-\partial _{\alpha }(\varepsilon _{0}\ ^{\ast
}F^{\alpha \beta })\gamma _{5}\gamma _{\beta }=c^{-1}j^{(C)\beta }\gamma
_{\beta },  \label{F8}
\end{equation}%
which can be separated into two equations, the equation with sources
\begin{equation}
\partial _{\alpha }(\varepsilon _{0}F^{\alpha \beta }+\mathcal{M}^{\alpha
\beta })\gamma _{\beta }=c^{-1}j^{(C)\beta }\gamma _{\beta }  \label{fs}
\end{equation}%
and the equation without sources, which is the same as in the vacuum
\begin{equation}
\partial _{\alpha }\ ^{\ast }F^{\alpha \beta }\gamma _{5}\gamma _{\beta }=0.
\label{fws}
\end{equation}%
In [31], the equation (\ref{F4}) with $F(x)$ and $\mathcal{M(}x\mathcal{)}$
is also written in terms of vectors $E$, $B$ and $P$, $M$. If written with $%
E $, $B$, $P$ and $M$ as CBGQs in the standard basis the equation with
sources (\ref{fs}) becomes
\begin{equation}
\partial _{\alpha }\{\varepsilon _{0}[\delta _{\quad \mu \nu }^{\alpha \beta
}E^{\mu }v^{\nu }+c\varepsilon ^{\alpha \beta \mu \nu }v_{\mu }B_{\nu
}]+[\delta _{\quad \mu \nu }^{\alpha \beta }P^{\mu }u^{\nu
}+(1/c)\varepsilon ^{\alpha \beta \mu \nu }M_{\mu }u_{\nu }]\}\gamma _{\beta
}=j^{(C)\beta }\gamma _{\beta },  \label{I1}
\end{equation}%
where $\delta _{\quad \mu \nu }^{\alpha \beta }=\delta _{\,\,\mu }^{\alpha
}\delta _{\,\,\nu }^{\beta }-\delta _{\,\,\nu }^{\alpha }\delta _{\,\,\mu
}^{\beta }$ and the equation without sources (\ref{fws}) becomes

\begin{equation}
\partial _{\alpha }(c\delta _{\quad \mu \nu }^{\alpha \beta }B^{\mu }v^{\nu
}+\varepsilon ^{\alpha \beta \mu \nu }E_{\mu }v_{\nu })\gamma _{5}\gamma
_{\beta }=0.  \label{I2}
\end{equation}%
In the ISR Eqs. (\ref{I1}) and (\ref{I2}) are fundamental equations for
moving media and \emph{they replace all usual Maxwell's equations (with
3-vectors) for moving media,} thus Eqs. (1-4) in [1] as well. Observe, as
pointed out in [31], that, in contrast to all conventional formulations of
the field equations for moving media, Eq. (\ref{I1}) contains two different
velocity vectors, $v$ - the velocity of the observers and $u$ - the velocity
of the moving medium, which come from the decompositions of $F$ and $%
\mathcal{M}$, Eqs. (\ref{fm}) and (\ref{mp}), respectively. Therefore, in
the general case, i.e., for $u\neq v$, it is not possible to introduce the
electric and magnetic excitations $D$ and $H$, respectively, where $%
D=\varepsilon _{0}E+P$ and $H=(1/\mu _{0})B-M$. The mentioned introduction
of $D$ and $H$ is possible if only one velocity, the velocity of the medium $%
u$, is taken into account, or the case $u=v$ is considered, or both
decompositions (\ref{fm}) and \ref{mp}) are made with the same velocity
vector, either $u$ or $v$, but that last case has not a proper physical
interpretation. This means that, in the general case $u\neq v$, Eqs. (1-4)
from [1] with 3-vectors $\mathbf{D}$ and $\mathbf{H}$ are not possible to
derive in a mathematically correct way from Eqs. (\ref{I1}) and (\ref{I2}).
All this is discussed in detail in Sections 6 and 7 in [31]. There, Eqs. (%
\ref{I1}) and (\ref{I2}) with 4D geometric quantities $E$, $B$, $P$ and $M$
are compared with the usual Maxwell equations with the 3-vectors $\mathbf{E}$%
, $\mathbf{B}$, $\mathbf{D}$, $\mathbf{H}$, which are the same as Eqs. (1-4)
in [1]. It is shown, as for the vacuum in [19], that in the 4D spacetime
Eqs. (\ref{I1}) and (\ref{I2}) \emph{are not equivalent }to the usual
Maxwell equations with the 3-vectors and their UT. The equations (\ref{I1})
and (\ref{I2}) hold for all relatively moving inertial observers and for all
bases used by them, which is not the case for, e.g., Eqs. (1-4) in [1].

Recently, in the same formulations with the 4D geometric quantities, the
constitutive relations and the magnetoelectric effect for moving media are
investigated in detail in [32].

The axiomatic geometric formulation of electromagnetism is also presented in
the modern textbook on classical electrodynamics [33], which uses the
calculus of exterior forms. The formulation from [33] deals with the
electromagnetic excitation tensor $\mathcal{H}$, which is decomposed into
the electric $D$ and magnetic excitations $H$ and with field equations for
them as the primary equations. As discussed above it is not correct in the
general case, $u\neq v$. Furthermore, Hehl and Obukhov, [33], introduce six
axioms (the charge conservation, the postulated expression for the Lorentz
force law, ..) and from them the field equations for $\mathcal{H}$ and $F$
are derived. It can be seen that all axioms from [33] can be derived from
only one axiom, the field equation for $F$. For vacuum, this is explicitly
shown in [12]. As pointed out in [31], the generalization to a moving medium
can be obtained simply replacing $F$ by $F+\mathcal{M}/\varepsilon _{0}$.
Some other ambiguities and shortcomings in the formulation from [33] are
discussed in detail in [32].\emph{\bigskip \bigskip }

\noindent \textit{5.2 The Lorentz force and momentum conservation
laws\bigskip }

It is also argued in [1, 5] that the usual Lorentz force is incompatible
with momentum conservation laws and has to be replaced by the Einstein-Laub
law; all with the 3-vectors. However, for the electromagnetic momentum that
correctly transforms under the LT see Sections 4 and 5 - 5.3 in [34] (only
components) and for the more general expressions with the 4D geometric
quantities see Section 2.6 in [12]. There, in Eqs. (37) - (43) in [12], a
basis - free expression for the most important quantity for the momentum and
energy of the electromagnetic field, the stress-energy vector $T(n)$, then
the expressions for the energy density $U$, the Poynting vector $S$ and the
momentum density $g$, the angular momentum density $M$ and the Lorentz force
$K_{L}$ are directly derived from the field equation for $F$ and they are
written exclusively in terms of $F$. The notation is as in [12].
Furthermore, the local conservation laws are also directly derived from that
field equation for $F$ and presented in Section 2.7 in [12], see Eqs. (48) -
(51). As mentioned in the preceding section, the generalization of these
relations to a moving medium is obtained replacing $F$ by $F+\mathcal{M}%
/\varepsilon _{0}$ and it is briefly discussed in Section 2 in [31]. The
Lorentz force law is completely compatible with momentum conservation laws,
but all quantities have to be the 4D geometric quantities.\emph{\bigskip }

\noindent \textit{5.3 \textquotedblleft The relativistic version of Newton's
law\textquotedblright\ \bigskip }

In [1], and also in the well-known textbooks, e.g., [10], [25-30], it is
also stated that the equation $\mathbf{F=}d\mathbf{p}/dt$ is
\textquotedblleft the relativistic version of Newton's
law.\textquotedblright\ However, as shown in [11] and also in [20], \emph{%
the equation}
\begin{equation}
\mathbf{F=}d\mathbf{p}/dt,\quad \mathbf{p=}m\gamma _{u}\mathbf{u}
\label{efp}
\end{equation}%
\emph{is not the relativistic equation of motion} since, contrary to the
common assertions, it is not covariant under the LT. \emph{Any 3D quantity
cannot correctly transform under the LT; it is not the same quantity for
relatively moving observers in the 4D spacetime.} Instead of the equation
with the 3D quantities one has to use the equation of motion with 4D
geometric quantities, Eq. (10) in [11],%
\begin{equation}
K=dp/d\tau ,\quad p=mu,  \label{lf}
\end{equation}%
where $p$ is the proper momentum vector and $\tau $ is the proper time, a
Lorentz scalar. In the 4D spacetime $p$, $\tau $ and $K$\ from (\ref{lf})
are the correctly defined quantities and not the 3D $\mathbf{p}$, the
coordinate time $t$ and the 3-force $\mathbf{F}$. The Lorentz force $K_{L}$
can be defined in terms of $F$, or using the decomposition of $F$\ (\ref{feb}%
) in terms of $E$ and $B$, as%
\begin{equation}
K_{L}=(q/c)F\cdot u,\quad K_{L}=(q/c)\left[ (1/c)E\wedge v+(IB)\cdot v\right]
\cdot u.  \label{fl}
\end{equation}%
If written as a CBGQ in the standard basis $K_{L}$ becomes%
\begin{equation}
K_{L}=(q/c)F^{\mu \nu }u_{\nu }\gamma _{\mu },\quad K_{L}=(q/c^{2})[(v^{\nu
}u_{\nu })E^{\mu }+\varepsilon ^{\lambda \mu \nu \rho }v_{\lambda }u_{\nu
}cB_{\rho }-(E^{\nu }u_{\nu })v^{\mu }]\gamma _{\mu }.  \label{lo}
\end{equation}%
Particularly, the Lorentz force ascribed by an observer comoving with a
charge, $u=v$, i.e., if the charge and the observer world lines coincide,
then $K_{L}$ is purely electric, $K_{L}=qE$. In Section 6.1 in [12], under
the title \textquotedblleft The Lorentz force and the motion of charged
particle in the electromagnetic field $F$\textquotedblright\ the definition
of $K_{L}$ in terms of $F$ is exclusively used ($K_{L}=(q/c)F\cdot u$)
without introducing the electric and magnetic fields. The quantities $K$ ($%
K_{L}$), $p$, $u$ transform in the same way, like any other vector, i.e.,
according to the LT (for the components in the standard basis they are the
same as the above mentioned LT of $E^{\mu }$ (\ref{el})) and not according
to the awkward UT of the 3-force $\mathbf{F}$, e.g., Eqs. (12.66) and
(12.67) in [25], and the 3-momentum $\mathbf{p}$, i.e., the 3-velocity $%
\mathbf{u}$.\bigskip

\noindent \textit{5.4 The charge densities in an infinite wire with a steady
current.}

\textit{Is magnetism }a relativistic phenomenon?\textit{\bigskip }

In [1], Mansuripur mentions and \textquotedblleft
resolves\textquotedblright\ yet another apparent conflict with relativity,
i.e., another paradox, that refers to the force on a current-carrying wire.
In order to show that in this case too there is no paradox we shall need to
know the charge densities for a moving or stationary infinite wire with a
steady current. They are considered, e.g., in the well-known textbooks
[25-30]. In [25], in Section 12.3.1 under the title \textquotedblleft
Magnetism as a Relativistic Phenomenon\textquotedblright\ it is assumed, as
in all other conventional treatments, that Clausius' hypothesis holds, i.e.,
that such \emph{stationary} wire with a steady current is globally and \emph{%
locally} charge neutral; $\rho ^{\prime }=\rho _{+}^{\prime }+\rho
_{-}^{\prime }=\rho _{0}^{\prime }-\rho _{0}^{\prime }=0$. $\rho
_{+}^{\prime }$\ is the charge density of the stationary ions, which is the
same as in that wire but without current, $\rho _{0}^{\prime }$, whereas $%
\rho _{-}^{\prime }$\ is the charge density of the \emph{moving} electrons
that is taken to be the same as in that charge neutral wire without current,
$-\rho _{0}^{\prime }$. In $S$, in which the wire is moving, in Section
12.3.1 in [25], it is argued: \textquotedblleft \textit{Conclusion}: As a
result of unequal \emph{Lorentz contraction} (my emphasis) of the positive
and negative lines, a current-carrying wire that is electrically neutral in
one inertial system will be charged in another.\textquotedblright\ The same
conclusion is obtained in, e.g., [26-30]. The net charge density of the
moving wire with steady current $\rho \neq 0$ sets up an \emph{external
electric field}. The essential point is that in the conventional formulation
of SR the charge density of the moving charges is considered to be a
well-defined quantity. Simply, it is increased by the Lorentz factor $\gamma
$ because of the Lorentz contraction of the moving length or volume and
because it is assumed that the charge defined by Eq. (\ref{qcl}) is the
Lorentz invariant charge. In the above example, e.g., $\rho _{+}=\gamma \rho
_{0}^{\prime }$. Purcell, Section 5.9 in [26] and Griffiths, Section 12.3.1
in [25], calculate the external electric field for the moving
current-carrying wire and they also determine the electrical force on the
test charge $q$ in the rest frame of that test charge; it is denoted as $%
\overline{S}$ in [25]. In that frame $\overline{S}$ the current-carrying
wire is moving; the wire is charged and $q$ is at rest. Then, they argue
that if there is a force on $q$ in one frame, the $\overline{S}$ frame,
there must be the force in the rest frame of the wire; it is denoted as $S$
in [25] and it is our $S^{\prime }$ frame. They calculate that \textit{non}%
electrical force in the rest frame of the wire, where the wire is supposed
to be neutral, using the UT of the force 3-vector, Eqs. (12.65)-(12.67) in
[25]. It is stated in [25]: \textquotedblleft Taken together, then,
electrostatics and relativity imply the existence of another force. This
\textquotedblleft other force\textquotedblright\ is, of course, \textit{%
magnetic}.\textquotedblright\ It is written in the form of Eq. (12.85) in
[25], which is the same as the usual expression for the Lorentz force
exerted by the magnetic field of a long, straight wire on a moving charge $q$%
. According to that result, the authors of [26] and [25], and also many
other authors of textbooks and papers who used the consideration from [26],
concluded that magnetism is a relativistic phenomenon. But, in the 4D
spacetime, such a conclusion is not relativistically correct. As explained
below, in the 4D spacetime, the Lorentz contraction has not well-defined
physical meaning. Furthermore, in [26] and [25] as in all other usual
treatments, the conventional definition of charge in terms of 3D quantities
is used in which the values of the charge density $\rho (\mathbf{r},t)$ are
taken simultaneously at some $t$ for all $\mathbf{r}$ in the 3D volume $V(t)$
over which $\rho $\ is integrated,
\begin{equation}
Q=\int_{V(t)}\rho (\mathbf{r},t)dV.  \label{qcl}
\end{equation}%
The same equation is supposed to be valid in some relatively moving inertial
frame of reference with primed quantities replacing the unprimed ones, see
Eqs. (50, 51) in Section 7.1 in [2] and references therein. But, observe
that $t^{\prime }$ in $S^{\prime }$ is not connected in any way with $t$ in $%
S$. Contrary to the generally accepted opinion, the charge $Q$\ defined in
such a way is not invariant under the LT. Instead of that usual definition
with 3D quantities we deal with the 4D quantities. The total electric charge
$Q$ in a three-dimensional hypersurface $H$ (with two-dimensional boundary $%
\delta H$) is defined as a Lorentz scalar by the equation%
\begin{equation}
Q_{\delta H}=(1/c)\int_{H}j\cdot ndH,  \label{qh}
\end{equation}%
where $j$ is the current density vector and the vector $n$ is the unit
normal to $H$.

Many years ago, it was shown by Rohrlich [35] and Gamba [36] that the
Lorentz contraction has nothing to do with the LT. It is, according to
Rohrlich [35], an \textquotedblleft apparent\textquotedblright\
transformation (AT) that does not refer to the same physical quantity in the
4D spacetime, whereas the transformations that refer to the same 4D physical
quantity were called the \textquotedblleft true
transformations\textquotedblright . The LT are the \textquotedblleft true
transformations\textquotedblright . Two spatial lengths that are
synchronously determined for the observers, the rest length and the Lorentz
contracted length, are two \emph{different} quantities in the 4D spacetime
and accordingly they cannot be connected by the LT, see Section 4.1 and Fig.
3 in [7] and compare it with the spacetime length, Section 3.1 and Fig. 1 in
[7], which is the same 4D quantity for all relatively moving inertial
observers. The LT do not connect two spatial lengths taken alone, i.e., in
the 4D spacetime, two relatively moving observers cannot compare spatial
lengths taken alone. Rohrlich's and Gamba's ideas are properly
mathematically formulated using 4D geometric quantities in [6-9]; for the
proof of the relativistic incorrectness of the Lorentz contraction see also
Appendix in [2]. In [7-9] it is proved that the time dilation is also an AT,
which has nothing in common with the LT and that both the Lorentz
contraction and the time dilation are not the intrinsic relativistic
effects. Note that the UT (\ref{be}) and the UT of $\mathbf{P}$ and $\mathbf{%
M}$, EDM $\mathbf{p}$ and MDM $\mathbf{m}$ are also the AT and, as shown in
[11], the same holds for the transformations of other 3D quantities, like
the AT of the 3D force, e.g., Eqs. (12.65)-(12.67) in [25], the AT of the 3D
torque (\ref{te}), the AT of the 3D angular momentum, e.g., in [16], etc.

In the 4D spacetime the properly defined 4D geometric quantities are the
position vectors $x_{A}$, $x_{B}$, of the events $A$ and $B$, respectively,
the distance vector $l_{AB}=x_{B}-x_{A}$ between two events and the
spacetime length $l$, $l=\mid l_{AB}^{\mu }l_{AB,\mu }\mid ^{1/2}$, which
is, e.g., for a moving rod, $l=L_{0}$, where $L_{0}$ is the rest length. The
spacetime length $l$ is a Lorentz scalar, see, e.g., [7].

As already mentioned, the LT cannot transform the spatial (temporal)
distance between two events again to the spatial (temporal) distance. Hence,
in the 4D spacetime, the Lorentz contracted length is meaningless and \emph{%
only} \emph{the rest length is a well-defined quantity.} In [6], an apparent
relativistic paradox is investigated both with the 4D geometric quantities
and with the conventional SR. That paradox is connected with the Lorentz
contraction and it often appears in different textbooks and papers under the
different names, e.g., in [37], and the same in [6], it is called
\textquotedblleft Car and garage paradox,\textquotedblright\ whereas in [25]
it is called \textquotedblleft the barn and ladder paradox\textquotedblright
, etc. It is shown in [6] that, in contrast to [37], [25], etc., i.e., to
the conventional formulation of SR, i.e., Einstein's formulation of SR,
there is no paradox in the formulation of SR with 4D geometric quantities,
i.e., in the ISR. In the Lorentz contraction the relatively moving observers
make the \emph{same measurements} (synchronously determine the spatial
length), \emph{but they do not look at the same 4D quantity, i.e., at the
same set of events.} On the other hand, as already stated, the LT refer to
the same 4D quantity, i.e., to the same set of events. It can be easily seen
from Section 4.1 and Fig. 3 in [7], or from Appendix in [2], that in the
Lorentz contraction the relatively moving observers do not look at the same
set of events. This means that the Lorentz contraction has nothing to do
with the LT, i.e., with the SR, which is \emph{the theory of the 4D
spacetime with the pseudo-Euclidean geometry.} Only the transformations that
leave the pseudo-Euclidean geometry of the 4D spacetime unchanged are the
relativistically correct transformations. The time dilation and the Lorentz
contraction are not such type of transformations, whereas the LT belong to
that category of transformations.

According to this discussion it is clear that the assertion from [26] and
[25] that magnetism is a relativistic phenomenon is meaningless in the 4D
spacetime. That conclusion is obtained using the Lorentz contraction and the
definition of charge in terms of 3D quantities (\ref{qcl}), which are not
well-defined in the 4D spacetime. Moreover, the axiomatic geometric
formulation from [12] explicitly shows that the electromagnetic field $F$ is
the primary quantity, which means that \emph{the whole electromagnetism is a
relativistic phenomenon.}

The fact that only the rest length is properly defined entails that \emph{in
the 4D spacetime it is not possible to give a definite physical meaning to
the charge density of moving charges. }In the 4D spacetime the usual charge
density $\rho $ and the usual current density $\mathbf{j}$ as a 3-vector are
not well-defined physical quantities, but it is only the current density
vector $j$, or as a CBGQ, e.g., in the standard basis, $j=j^{\mu }\gamma
_{\mu }$. Hence, as shown in Sections 3 and 3.1 in [6], or in Section 7.1 in
[2], in order to determine the current density vector $j^{\mu }$ in some
inertial frame of reference in which the charges are moving we first have to
determine that vector in the rest frame of the charges, where the spatial
components $j^{i}$ are zero and only the temporal component $j^{0}\neq 0$
and then to transform by the LT so determined $j^{\mu }\gamma _{\mu }$ to
the considered inertial frame of reference. This means that in order to
determine the current density vector $j$ in some inertial frame of reference
for an infinite wire with a steady current we first have to determine the
current density vectors $j_{+}^{\mu }\gamma _{\mu }$ and $j_{-}^{\mu }\gamma
_{\mu }$ for positive and negative charges, respectively, in their rest
frames. Then, they have to be transformed by the LT to the given inertial
frame of reference. Thereby, in the rest frame of the wire, the $S^{\prime }$
frame, the positive charges are at rest and $j_{+}$ as a CBGQ is
\begin{equation}
j_{+}=j_{+}^{\prime \mu }\gamma _{\mu }^{\prime }=(c\rho _{0}^{\prime
})\gamma _{0}^{\prime }+0\gamma _{1}^{\prime }.  \label{jpl}
\end{equation}%
The negative charges are moving in $S^{\prime }$ and, according to the above
discussion, we first have to write the current density vector of the
electrons in the frame, let us denote it as $S_{e}$, in which the spatial
components of the vector $j_{-}$ ($j_{-}=j_{e,-}^{\mu }\gamma _{e,\mu }$; $%
j_{e,-}^{\mu }$ are the components of $j_{-}$\ in the standard basis and in
the $S_{e}$ frame, whereas $\gamma _{e,\mu }$\ are the unit vectors in $%
S_{e} $) are zero, $j_{e,-}^{i}=0$. In the usual notation, in $S_{e}$, the
drift velocity 3-vector of the electrons is zero. Hence, the temporal
component $j_{e,-}^{0}$\ in $S_{e}$ is a well-defined quantity. Remember
that the electrons, in average, are not moving in $S_{e}$, which means that
the situation for the electrons is the same as in that wire but without any
current, i.e., $j_{e,-}^{0}=c\rho _{e,-}=-c\rho _{0}^{\prime }$; $\rho
_{e,-} $ is the same as the proper charge density of the electrons $-\rho
_{0}^{\prime }$. Observe that such result is completely different than the
Clausius hypothesis. The current density vector of the electrons $j_{-}$ as
a CBGQ in $S_{e}$ is%
\begin{equation}
j_{-}=j_{e,-}^{\mu }\gamma _{e,\mu }=(-c\rho _{0}^{\prime })\gamma
_{e,0}+0\gamma _{e,1}.  \label{ronula}
\end{equation}%
Then, by means of (\ref{ronula}) and the LT one finds the current density
vector of the electrons in $S^{\prime }$, the rest frame of the wire, i.e.,
the lab frame, as%
\begin{equation}
j_{-}=j_{-}^{\prime \mu }\gamma _{\mu }^{\prime }=(-c\gamma _{e}\rho
_{0}^{\prime })\gamma _{0}^{\prime }+(-c\gamma _{e}\beta _{e}\rho
_{0}^{\prime })\gamma _{1}^{\prime },  \label{jotmi}
\end{equation}%
where $\gamma _{e}=(1-\beta _{e}^{2})^{-1/2}$ and $\beta _{e}=v_{d}/c$, $%
v_{d}$ is the usual drift velocity of the electrons in that stationary wire
with current. Similarly, $j_{+}$ as a CBGQ in $S_{e}$ can be determined
using the LT of quantities from (\ref{jpl}). This yields that $j_{+}$ in $%
S_{e}$ is%
\begin{equation}
j_{+}=j_{e,+}^{\mu }\gamma _{e,\mu }=(c\gamma _{e}\rho _{0}^{\prime })\gamma
_{e,0}+(-c\gamma _{e}\beta _{e}\rho _{0}^{\prime })\gamma _{e,1}  \label{jme}
\end{equation}%
It can be seen from (\ref{jpl}) and (\ref{jme}) that $j_{+}=j_{+}^{\prime
\mu }\gamma _{\mu }^{\prime }=j_{e,+}^{\mu }\gamma _{e,\mu }$ and the same
for $j_{-}$\ using (\ref{ronula}) and (\ref{jotmi}). The total current
density vector in $S^{\prime }$ is $j=j^{\prime \mu }\gamma _{\mu }^{\prime
} $, where the components in the standard basis are $j^{\prime \mu
}=j_{+}^{\prime \mu }+j_{-}^{\prime \mu }$. As we know, $j_{+}^{\prime \mu
}=(c\rho _{0}^{\prime },0)$ and $j_{-}^{\prime \mu }$ are given by (\ref%
{jotmi}), which yields that $j$ as a CBGQ in $S^{\prime }$, the rest frame
of the wire, is%
\begin{equation}
j=j^{\prime \mu }\gamma _{\mu }^{\prime }=c(1-\gamma _{e})\rho _{0}^{\prime
}\gamma _{0}^{\prime }+(-c\gamma _{e}\beta _{e}\rho _{0}^{\prime })\gamma
_{1}^{\prime };  \label{swj}
\end{equation}%
the temporal component $j^{\prime 0}$ \emph{is not zero.} Observe that it
will again hold that $j=j^{\prime \mu }\gamma _{\mu }^{\prime }=j_{e}^{\mu
}\gamma _{e,\mu }$, where the quantities in $S^{\prime }$ and in $S_{e}$ are
connected by the LT. Then, the result (\ref{swj}) causes that, in contrast
to the usual approaches, there is an \emph{external electric field} not only
outside moving wire with a steady current but also outside that stationary
wire. All this, together with the expression for the \emph{external electric
field} from a \emph{stationary} wire with a steady current, is already
discussed in a slightly different way in Sections 3 - 3.3 in [6] and 7.1 in
[2].

An infinite wire is not a physical system and therefore the above results
are generalized to a current loop in Sections 4 in [6] and discussed also in
Section 7.1 in [2]. It is shown there that the external electric field
exists not only for a moving current loop, as in the usual approaches, but
\emph{for the stationary current loop as well}. Such a current loop, moving
or stationary, always behaves at points far from that current loop like an
electric dipole, but as a 4D geometric quantity. In Section 7.2 in [2]
different experiments for the detection of that dipole electric field from a
stationary current loop are discussed.\emph{\bigskip }

\noindent \textit{5.5 Is there a conflict with relativity for the force on a
current-carrying wire\bigskip }

Having determined the charge densities in a current-carrying wire we turn to
the investigation of the force on such a current-carrying wire. Mansuripur
[1] considers \textquotedblleft a thin, straight, charge-neutral wire
carrying a constant, uniform current density $\mathbf{J}_{\text{free}}$
along $x^{\prime }$ in the presence of a constant, uniform \textit{E} field
(also along $x^{\prime }$)\textquotedblright\ Remember that in [1] the $%
S^{\prime }$ frame moves along the $z$ axis, as seen by a stationary
observer in $S$. It is argued in [1] that in its rest frame (the $S^{\prime
} $ frame) the wire does not experience a Lorentz force, but if seen by the
stationary observer in the $S$ frame it does experience a Lorentz force
along the $z$ axis. Let us try to explain how this result is obtained. In $%
S^{\prime }$ the Lorentz force density
\begin{equation}
\mathbf{f}^{\prime }=\rho ^{\prime }\mathbf{E}^{\prime }+\mathbf{J}^{\prime
}\times \mathbf{B}^{\prime }  \label{foc}
\end{equation}%
is zero because, for a stationary wire with a steady current it is assumed
that $\rho ^{\prime }=0$ (Clausius' hypothesis, see, as mentioned above,
Sections 3 and 3.1 in [6] or 7.1 and 7.2 in [2] and references therein) and $%
\mathbf{B}^{\prime }=\mathbf{0}$. In $S$, the components $j^{\mu }$ are $%
\rho =0$, $j_{x}=j_{x}^{\prime }$, $j_{y}=j_{z}=0$. According to the UT (\ref%
{be}) the components of the 3-vector $\mathbf{E}$ are $E_{x}=\gamma
E_{x}^{\prime }$, $E_{y}=E_{z}=0$ and \emph{there is an induced component of
the magnetic field 3-vector}, i.e., $B_{y}=-\gamma \beta E_{x}^{\prime }$, $%
B_{x}=B_{z}=0$, which yields that there is $f_{z}=j_{x}B_{y}\neq 0$. It is
clear that again the real cause of the existence of the paradox in that
example is the use of the 3-vectors and their UT. How does Mansuripur
\textquotedblleft resolve\textquotedblright\ this paradox? He argues
\textquotedblleft .. special relativity is not violated here because the
energy delivered by the \textit{E} field at the rate of $\mathbf{E}\cdot
\mathbf{J}_{free}$ to the current causes \emph{an increase in the mass of
the wire}. (my emphasis) Seen by the observer in the $xyz$ frame, the wire
has a nonzero (albeit constant) velocity along $z$, and, therefore, its
relativistic momentum $\mathbf{p}$ increases with time, not because of a
change of velocity but because of a change of mass. The observed
electromagnetic force in the moving frame thus agrees with the relativistic
version of Newton's law.\textquotedblright\ From the point of view of the
ISR such a \textquotedblleft resolution\textquotedblright\ contains a wealth
of relativistically incorrect quantities and explanations. Again, the
\textquotedblleft resolution\textquotedblright\ exclusively deals with the
3-vectors, $\mathbf{E}$, $\mathbf{J}$, $\mathbf{p}$ and their UT.
Furthermore, in the 4D spacetime only the rest mass is well defined quantity
and thus \emph{there is not \textquotedblleft a change of
mass.\textquotedblright }\ Also, as explained above, $\mathbf{F=}d\mathbf{p}%
/dt$ \emph{is not the relativistic equation of motion}.

Barnett [4], in his Comment on [1], uses the similar argument as above but
for the \textquotedblleft resolution\textquotedblright\ of Mansuripur's
paradox. He declares: \textquotedblleft In Mansuripur's magnetic-dipole
thought experiment there is no change in the velocity of the dipole because
there is no net force acting on it, but \emph{there is a change in the
moment of inertia and this balances exactly the torque derived from }(1)%
\emph{.}\textquotedblright\ (my emphasis) The equation (1) in Barnett's
paper is the usual expression for the Lorentz force density, $\mathbf{f}%
=\rho \mathbf{E}+\mathbf{J}\times \mathbf{B}$. The torque obtained in
Barnett's paper is the 3D torque and it arises, as in all usual approaches
including all papers [1, 3-5], from the use of the UT of the 3D quantities
according to which, Barnett [4]: \textquotedblleft The moving magnetic
dipole, moreover, acquires some electric dipole character by virtue of its
motion.\textquotedblright\ In order to cancel this offending torque he
argues that the time component of the force as a \textit{four-vector}
\textquotedblleft the $\mathbf{J}\cdot \mathbf{E}$ term produces a change to
the moment of inertia of the dipole, .. .\textquotedblright\ All objections
that we have presented above hold in the same measure for Barnett's
\textquotedblleft resolution\textquotedblright\ of Mansuripur's paradox. In
the 4D spacetime there are no 3D quantities, the components (even in the
standard basis) of some 4-vector, i.e., vector on the 4D spacetime, cannot
be written in terms of the 3-vectors, the UT are not the LT, there is no
change to the moment of inertia, etc. Barnett, as in almost all usual
covariant approaches, e.g., [10, 25 - 30], considers that components
implicitly taken in the standard basis are, e.g., four-vector, or, more
generally tensors. \emph{Components} are numbers depending on the chosen
basis and mathematically they \emph{are not tensors}. It has to be pointed
out once again that in the 4D spacetime there are no 3D force $\mathbf{F}$,
3D magnetization $\mathbf{M}$, 3D torque $\mathbf{T}$, 3D acceleration $%
\mathbf{a}$, etc. The LT cannot transform the 3D quantities. They are
transformations of the 4D geometric quantities that are properly defined on
the 4D spacetime.

Now, let us examine the force on a current-carrying wire for the case
considered in [1] if it is treated with 4D geometric quantities. The Lorentz
force density $k_{L}$ as a correctly defined abstract vector is $%
k_{L}=(1/c)F\cdot j$, or as a CBGQ it is given by (\ref{LoF}). Inserting the
decomposition of $F$ (\ref{fm}) into (\ref{LoF}) $k_{L}$ becomes%
\begin{equation}
k_{L}=(1/c^{2})[(v^{\nu }j_{\nu })E^{\mu }+\varepsilon ^{\lambda \mu \nu
\rho }v_{\lambda }j_{\nu }cB_{\rho }-(E^{\nu }j_{\nu })v^{\mu }]\gamma _{\mu
}.  \label{kel}
\end{equation}%
That expression is correct in all bases and all quantities in (\ref{kel})
properly transform under the LT. $k_{L}$ from (\ref{kel}) replaces the usual
expression with the 3D quantities, like, e.g., (\ref{foc}). In $S^{\prime }$%
, the rest frame of the wire, which is taken to be the $\gamma _{0}$-frame, $%
v^{\prime \mu }=(c,0,0,0)$ and $E^{\prime \mu }=(0,E^{\prime 1},0,0)$ and
the components of current density vector are $j^{\prime \mu }=(j^{\prime
0},j^{\prime 1},0,0)$, where $j^{\prime 0}$ and $j^{\prime 1}$ are given by (%
\ref{swj}), $j^{\prime 0}=c(1-\gamma _{e})\rho _{0}^{\prime }$, $j^{\prime
1}=-c\gamma _{e}\beta _{e}\rho _{0}^{\prime }$. The magnetic field vector is
zero, $B=B^{\prime \mu }\gamma _{\mu }^{\prime }=0$ and it remains zero in
all relatively moving inertial frames of reference. Inserting these
components into (\ref{kel}) it is obtained that
\begin{equation}
k_{L}=k_{L}^{\prime \mu }\gamma _{\mu }^{\prime }=(1/c^{2})[-cE^{\prime
1}j_{1}^{\prime }\gamma _{0}^{\prime }+cE^{\prime 1}j_{0}^{\prime }\gamma
_{1}^{\prime }+0\gamma _{2}^{\prime }+0\gamma _{3}^{\prime }].  \label{kc}
\end{equation}%
In $S^{\prime }$, the Lorentz force density is not zero; there are both the
temporal component and a spatial component. This result is essentially
different than in [1]. All quantities in (\ref{kc}) are properly defined
quantities on the 4D spacetime, which correctly transform under the LT. $%
k_{L}$ given by (\ref{kel}), as a CBGQ, is an invariant quantity, i.e., it
will be the same as in (\ref{kc}) for all relatively moving inertial frames
of reference, thus for the $S$ frame from [1] as well, $k_{L}^{\prime \mu
}\gamma _{\mu }^{\prime }=k_{L}^{\mu }\gamma _{\mu }$. Let us explicitly
determine $k_{L}$ in the $S$ frame from [1]. This can be made in different
ways, e.g., by the LT of all quantities in (\ref{kc}) from $S^{\prime }$ to $%
S$, or by the LT of the Lorentz force density vector, i.e., of $%
k_{L}^{\prime \mu }$ to $k_{L}^{\mu }$ and $\gamma _{\mu }^{\prime }$ to $%
\gamma _{\mu }$, or to simply introduce into (\ref{kel}) the quantities from
the $S$ frame. We shall use the third possibility. In $S$, the
\textquotedblleft fiducial\textquotedblright\ observers are moving and $%
v^{\mu }=(\gamma c,0,0,\beta \gamma c)$. In the same way we find that $%
E^{\mu }=(0,E^{1}=E^{\prime 1},0,0)$ and $j^{\mu }=(j^{0}=\gamma j^{\prime
0},\ j^{1}=j^{\prime 1},0,j^{3}=\beta \gamma j^{\prime 0})$, and of course, $%
B^{\mu }=(0,0,0,0)$. Note that the electric field vector transforms by the
LT, as in (\ref{el}), again to the electric field vector and the same for
the magnetic field vector. This yields that
\begin{equation}
k_{L}=k_{L}^{\mu }\gamma _{\mu }=(1/c^{2})[-c\gamma E^{1}j_{1}\gamma
_{0}+E^{1}c\gamma (j_{0}+\beta j_{3})\gamma _{1}+0\gamma _{2}-c\beta \gamma
E^{1}j_{1}\gamma _{3}],  \label{ks}
\end{equation}%
where $\gamma =(1-\beta ^{2})^{-1/2}$ and $\beta =V/c$. It can be easily
shown that $k_{L}^{\mu }\gamma _{\mu }$ from (\ref{ks}) is $=k_{L}^{\prime
\mu }\gamma _{\mu }^{\prime }$ from (\ref{kc}); \emph{there is no paradox}
and there is no need for \textquotedblleft a change of
mass.\textquotedblright

Let us suppose for a moment that $j^{\prime 0}=0$, as in the usual
approaches. Then, from (\ref{kc}) it follows that $k_{L}=k_{L}^{\prime \mu
}\gamma _{\mu }^{\prime }=(-1/c)(E^{\prime 1}j_{1}^{\prime })\gamma
_{0}^{\prime }$. It corresponds to $\mathbf{J}\cdot \mathbf{E}$ term from
Barnett's paper [4], but, remember, that we deal with correctly defined
quantities in the 4D spacetime and with their LT and not with the 3D
quantities and their UT, i.e., the AT. Hence, in $S$, the LT of $%
k_{L}^{\prime \mu }\gamma _{\mu }^{\prime }$ give that $k_{L}=k_{L}^{\mu
}\gamma _{\mu }=(-1/c)[\gamma E^{1}j_{1}\gamma _{0}+\beta \gamma
E^{1}j_{1}\gamma _{3}]$ and again it holds $k_{L}=k_{L}^{\mu }\gamma _{\mu
}=k_{L}^{\prime \mu }\gamma _{\mu }^{\prime }$; it is the same quantity for
relatively moving inertial observers and \emph{there is no paradox}. It is
visible that $k_{L}$\ again contains the temporal component as in $S^{\prime
}$, but also a spatial component, which is in the $\gamma _{3}$\ direction.
Observe that the magnetic field vector is again zero both in $S^{\prime }$%
and $S$ as in (\ref{kc}) and (\ref{ks}), $B=B^{\prime \mu }\gamma _{\mu
}^{\prime }=B^{\mu }\gamma _{\mu }=0$, but, nevertheless, there is $\gamma
_{3}$\ component.\bigskip

\noindent \textit{5.6 The electromagnetic field of a point charge in uniform
motion\bigskip }

\noindent \textit{5.6.1 The bivector }$F$ \textit{for} \textit{an uniformly
moving charge\bigskip }

Furthermore, as already stated, in this geometric approach to SR, i.e., in
the ISR, the electromagnetic field $F$ yields the complete description of
the electromagnetic field and, in principle, there is no need to introduce
either the field vectors $E$ and $B$ or the 4D potential $A$. For the given
sources Eq. (\ref{df}), or Eq. (\ref{c1}), can be solved to give the
electromagnetic field $F$. The expression for $F$ for an arbitrary motion of
a point charge is given in [12] by Eq. (11) and, particularly, by Eq. (12)
for a charge $Q$ moving with constant velocity vector $u_{Q}$. It is%
\begin{equation}
F(x)=G(x\wedge (u_{Q}/c)),\quad G=kQ/\left\vert x\wedge (u_{Q}/c)\right\vert
^{3},  \label{cvf}
\end{equation}%
where $k=1/4\pi \varepsilon _{0}$. $G$ is a number, a Lorentz scalar,
whereas the geometric character of $F$ is contained in $x\wedge (u_{Q}/c)$. $%
F$ from (\ref{cvf}) can be written as a CBGQ in the standard basis, $%
F=(1/2)F^{\mu \nu }\gamma _{\mu }\wedge \gamma _{\nu }$,
\begin{equation}
F=G(1/c)x^{\mu }u_{Q}^{\nu }(\gamma _{\mu }\wedge \gamma _{\nu }),\quad
G=kQ[(x^{\mu }u_{Q,\mu })^{2}-c^{2}x^{\mu }x_{\mu }]^{3/2}.  \label{cf}
\end{equation}%
The basis components $F^{\mu \nu }$ are determined as $F^{\mu \nu }=\gamma
^{\nu }\cdot (\gamma ^{\mu }\cdot F)=(\gamma ^{\nu }\wedge \gamma ^{\mu
})\cdot F$,%
\begin{equation}
F^{\mu \nu }=G(1/c)(x^{\mu }u_{Q}^{\nu }-x^{\nu }u_{Q}^{\mu }).  \label{fmn}
\end{equation}%
In Section 4, for simplicity, we have dealt with a uniform electric field.
However, as already discussed at the beginning of Section 4, in the ISR it
would be more appropriate to exclusively deal with the primary quantity $F$,
i.e., $F^{\mu \nu }$.

Let us write the expression for $F$ (\ref{cf}) in the $S^{\prime }$ frame in
which the charge $Q$ is at rest, i.e., $u_{Q}/c=\gamma _{0}^{\prime }$ with $%
\gamma _{0}^{\prime \mu }=(1,0,0,0)$. Then, $F=(1/2)F^{\prime \mu \nu
}\gamma _{\mu }^{\prime }\wedge \gamma _{\nu }^{\prime }$ and
\begin{equation}
F=F^{\prime i0}(\gamma _{i}^{\prime }\wedge \gamma _{0}^{\prime
})=Gx^{\prime i}(\gamma _{i}^{\prime }\wedge \gamma _{0}^{\prime }),\quad
G=kQ/(x^{\prime i}x_{i}^{\prime })^{3/2}.  \label{q1}
\end{equation}%
In $S^{\prime }$ and in the standard basis, the basis components $F^{\prime
\mu \nu }$ of the bivector $F$ are obtained from (\ref{fmn}) and they are:
\begin{equation}
F^{\prime i0}=-F^{\prime 0i}=kQx^{\prime i}/(x^{\prime i}x_{i}^{\prime
})^{3/2},\quad F^{\prime ij}=0.  \label{fci}
\end{equation}%
In the same way we find the expression for $F$ (\ref{cf}) in the $S$ frame
in which the charge $Q$ is moving, i.e., $u_{Q}=u_{Q}^{\mu }\gamma _{\mu }$
with $u_{Q}^{\mu }/c=(\gamma _{Q},\gamma _{Q}\beta _{Q},0,0)$. Then
\begin{gather}
F=G\gamma _{Q}[(x^{1}-\beta _{Q}x^{0})(\gamma _{1}\wedge \gamma
_{0})+x^{2}(\gamma _{2}\wedge \gamma _{0})+x^{3}(\gamma _{3}\wedge \gamma
_{0})  \notag \\
-\beta _{Q}x^{2}(\gamma _{1}\wedge \gamma _{2})-\beta _{Q}x^{3}(\gamma
_{1}\wedge \gamma _{3})],\quad G=kQ/[\gamma _{Q}^{2}(x^{1}-\beta
_{Q}x^{0})^{2}+(x^{2})^{2}+(x^{3})^{2}]^{3/2}.  \label{fsq}
\end{gather}%
In $S$ and in the standard basis, the basis components $F^{\mu \nu }$ of the
bivector $F$ are again obtained from (\ref{fmn}) and they are%
\begin{eqnarray}
F^{10} &=&G\gamma _{Q}[(x^{1}-\beta _{Q}x^{0}),\ F^{20}=G\gamma _{Q}x^{2},\
F^{30}=G\gamma _{Q}x^{3},  \notag \\
F^{21} &=&G\gamma _{Q}\beta _{Q}x^{2},\ F^{31}=G\gamma _{Q}\beta _{Q}x^{3},\
F^{32}=0.  \label{fi0}
\end{eqnarray}%
The expression for $F$ as a CBGQ in the $S$ frame can be find in another way
as well, i.e., to make the LT of the quantities from Eq. (\ref{q1}). Observe
that the CBGQs from (\ref{q1}) and (\ref{fsq}), which are the
representations of the bivector $F$ in $S^{\prime }$ and $S$ respectively,
are equal, $F$ from (\ref{q1}) $=$ $F$ from (\ref{fsq}); \emph{they are the
same quantity }$F$ \emph{from} (\ref{cf}) \emph{for observers in} $S^{\prime
}$ \emph{and} $S$.

In Section 2 it is explained that in the usual approaches to the
relativistic electrodynamics the components of the 3-vectors $\mathbf{E}$
and $\mathbf{B}$ in $S^{\prime }$ and $S$ are identified with the components
$F^{\alpha \beta }$ implicitly taken in the standard basis and that the UT
of the components of $\mathbf{E}$ and $\mathbf{B}$ are derived assuming that
they transform under the LT as the components $F^{\alpha \beta }$ transform.
Then, the 3-vectors $\mathbf{E}^{\prime }$, $\mathbf{B}^{\prime }$ in $%
S^{\prime }$ and $\mathbf{E}$, $\mathbf{B}$ in $S$ are constructed in the
same way, i.e. multiplying the components $E_{x,y,z}^{\prime }$, $%
B_{x,y,z}^{\prime }$ by the unit 3-vectors $\mathbf{i}^{\prime }$, $\mathbf{j%
}^{\prime }$, $\mathbf{k}^{\prime }$ and the components $E_{x,y,z}$, $%
B_{x,y,z}$ by the unit 3-vectors $\mathbf{i}$, $\mathbf{j}$, $\mathbf{k}$,
respectively. Such a procedure yields the UT of the 3-vectors $\mathbf{E}$
and $\mathbf{B}$\textbf{, }Eq. (11.149) in [10]. The mathematical
incorrectness of that procedure can be nicely illustrated comparing Eqs. (%
\ref{q1}) and (\ref{fci}) with Eq. (11) from [1] and Eqs. (\ref{fsq}) and (%
\ref{fi0}) with Eqs. (12a) and (12b) in [1].

The equations (\ref{q1}) and (\ref{fsq}) reveal that the physical quantities
are not only the components (\ref{fci}) and (\ref{fi0}), respectively, but
\emph{these components multiplied by the bivector bases in} $S^{\prime }$
\emph{and} $S$. Only if components and bases are taken together like in (\ref%
{q1}) and (\ref{fsq}), \emph{these CBGQs represent the same quantity}, $F$
from (\ref{cvf}). In the 4D spacetime, all quantities in (\ref{q1}) and (\ref%
{fsq}) are correctly defined and they properly transform under the LT; the
CBGQ from (\ref{fsq}) is the LT of the CBGQ from (\ref{q1}).

The situation is completely different in Eqs. (11), (12a) and (12b) in [1].
The components of the 3-vector $\mathbf{E}^{\prime }$ in Eq. (11) in [1] are
the same as the components in (\ref{fci}), but they are multiplied by the
unit 3-vectors $\mathbf{i}^{\prime }$, $\mathbf{j}^{\prime }$, $\mathbf{k}%
^{\prime }$; $\mathbf{E}^{\prime }$ \emph{is a geometric quantity in the 3D
space.} Mathematically, this is an incorrect step; the components of the 4D
geometric quantity are multiplied by the unit vectors from the 3D space to
form a 3D vector. Similarly, the components of the 3-vectors $\mathbf{E}$
and $\mathbf{B}$ in Eqs. (12a) and (12b) in [1] are the same as the
components in (\ref{fi0}), but the same remarks about the mathematical
incorrectness of the construction of $\mathbf{E}$ and $\mathbf{B}$ hold in
this case as well. It is stated in [1]: \textquotedblleft When the above E
field ($\mathbf{E}^{\prime }$ in Eq. (11), my remark) is Lorentz transformed
to the xyz frame, the resulting fields will be ($\mathbf{E}$ and $\mathbf{H}$
in Eqs. (12a) and (12b), my remark).\textquotedblright\ The LT always act on
the 4D spacetime and consequently they cannot transform the unit 3-vectors $%
\mathbf{i}^{\prime }$, $\mathbf{j}^{\prime }$, $\mathbf{k}^{\prime }$ into
the unit 3-vectors $\mathbf{i}$, $\mathbf{j}$, $\mathbf{k}$. Moreover, there
is not any kind of transformations which transform the 3-vectors from one 4D
frame to the 3-vectors from relatively moving 4D frame. Furthermore, as
explained at the end of Section 2, if instead of the standard basis the
observers use the $\left\{ r_{\mu }\right\} $ basis with the
\textquotedblleft radio\textquotedblright\ synchronization then the
space-time split of the 4D spacetime is not possible and the identification
of the components of $\mathbf{E}$ and $\mathbf{B}$ with the components $%
F^{\alpha \beta }$ is meaningless. This consideration, once again,
explicitly shows that in the 4D spacetime there is no room for the 3-vectors
and accordingly that the transformations that transform the $\mathbf{E}%
^{\prime }$ field given by Eq. (11) into $\mathbf{E}$ and $\mathbf{H}$
fields given by Eqs. (12a) and (12b) in [1] have nothing in common with the
relativistically correct LT as the transformations that are defined on the
4D spacetime.

In the 4D spacetime it is appropriate to deal with the abstract $F$ defined
by (\ref{cvf}), or with its representations, the CBGQs, defined by (\ref{q1}%
) and (\ref{fsq}), but not with $\mathbf{E}^{\prime }$ and $\mathbf{E}$, $%
\mathbf{H}$ that are defined by Eq. (11) and by Eqs. (12a), (12b) in [1],
respectively.\bigskip

\noindent \textit{5.6.2 The }vectors $E$ \textit{and} $B$ \textit{for a
charge} $Q$ \textit{moving with constant velocity} $u_{Q}$\textit{\bigskip }

Instead of to deal exclusively with $F$ we can construct in a mathematically
correct way vectors $E$ and $B$ for a charge $Q$ moving with constant
velocity $u_{Q}$. \emph{The vectors} $E$ \emph{and} $B$ are \emph{explicitly}
\emph{observer dependent}, i.e., dependent on $v$. For the same $F$ the
vectors $E$ and $B$ will have different expressions depending on the
velocity of observers who measure them. Using (\ref{eba}) and $F$ from (\ref%
{cvf}) we find the expressions for $E$ and $B$ in the form%
\begin{eqnarray}
E &=&(G/c^{2})[(u_{Q}\cdot v)x-(x\cdot v)u_{Q}],  \notag \\
B &=&(-G/c^{3})I(x\wedge u_{Q}\wedge v).  \label{ebe}
\end{eqnarray}%
It is worth mentioning that $E$ and $B$ from (\ref{ebe}) depend on two
velocity vectors $u_{Q}$ and $v$, whereas the 3-vectors $\mathbf{E}$ and $%
\mathbf{B}$ depend only on the 3-velocity of the charge $Q$. If the world
lines of the observer and the charge $Q$ coincide, $u_{Q}=v$, then (\ref{ebe}%
) yields that $B=0$ and only an electric field (Coulomb field) remains. It
can be seen that if $E$ and $B$ from (\ref{ebe}) are introduced into $F$
from (\ref{feb}) then they will yield $F$ defined by (\ref{cvf}), which
contains only $u_{Q}$, the velocity of the charge $Q$\ and not the velocity
of the observer $v$. This result directly proves that \emph{the
electromagnetic field} $F$ \emph{is the primary quantity from which the
observer dependent} $E$ \emph{and} $B$ \emph{are derived.}

If the CBGQs are used then the expressions for $E$ and $B$, Eq. (\ref{ebv}),
and that one for $F$ (\ref{cf}) yield $E$ and $B$, Eq. (\ref{ebe}), written
as CBGQs in the standard basis%
\begin{eqnarray}
E &=&E^{\mu }\gamma _{\mu }=(G/c^{2})[(u_{Q}^{\nu }v_{\nu })x^{\mu }-(x^{\nu
}v_{\nu })u_{Q}^{\mu }]\gamma _{\mu },  \notag \\
B &=&B^{\mu }\gamma _{\mu }=(G/c^{3})\varepsilon ^{\mu \nu \alpha \beta
}x_{\nu }u_{\alpha }v_{\beta }\gamma _{\mu }.  \label{ecb}
\end{eqnarray}%
If $E$ and $B$ from (\ref{ecb}) are introduced into $F$ from (\ref{fm}) then
they will yield $F$ as the CBGQ that is defined by (\ref{cf}). Again,
although $E$ and $B$ as the CBGQs from (\ref{ecb}) depend not only on $u_{Q}$
but on $v$ as well the electromagnetic field $F$ from (\ref{cf}) does not
contain the velocity of the observer $v$.\bigskip

\noindent \textit{5.6.2.1 The \textquotedblleft fiducial\textquotedblright\
observers are in }$S^{\prime }$, $v=c\gamma _{0}^{\prime }$, \textit{which
is the rest frame of the charge} $Q$\textit{\bigskip }

Let us take that the observers who measure $E$, $B$ fields are at rest, $%
v=c\gamma _{0}^{\prime }$, in the rest frame of the charge $Q$, $%
u_{Q}=v=c\gamma _{0}^{\prime }$. This means that the $S^{\prime }$ frame is
the $\gamma _{0}$-frame; the \textquotedblleft fiducial\textquotedblright\
observers with the $\left\{ \gamma _{\mu }\right\} $ basis. It follows from (%
\ref{ecb}) that
\begin{equation}
E=E^{\prime \mu }\gamma _{\mu }^{\prime }=Gx^{\prime i}\gamma _{i}^{\prime
},\quad E^{\prime 0}=0,\quad G=kQ/(x^{\prime i}x_{i}^{\prime })^{3/2};\quad
B=B^{\prime \mu }\gamma _{\mu }^{\prime }=0.  \label{bf}
\end{equation}%
This result agrees with the usual result, e.g., Eq. (11) in [1]. Now comes
the essential difference relative to all usual approaches. In order to find
the representations of $E$ and $B$ in $S$, i.e., the CBGQs $E^{\mu }\gamma
_{\mu }$ and $B^{\mu }\gamma _{\mu }$, we can either perform the LT of $%
E^{\prime \mu }\gamma _{\mu }^{\prime }$ and $B^{\prime \mu }\gamma _{\mu
}^{\prime }$ that are given by (\ref{bf}), or simply to take in (\ref{ecb})
that \emph{both} the charge $Q$ \emph{and the \textquotedblleft
fiducial\textquotedblright\ observers} are moving relative to the observers
in $S$; $v^{\mu }=u_{Q}^{\mu }=(\gamma _{Q}c,\beta _{Q}\gamma _{Q}c,0,0)$.
This yields the CBGQs $E^{\mu }\gamma _{\mu }$ and $B^{\mu }\gamma _{\mu }$
in $S$ with the condition that the \textquotedblleft
fiducial\textquotedblright\ observers are in $S^{\prime }$, $v=c\gamma
_{0}^{\prime }$, which is the rest frame of the charge $Q$, $u_{Q}=c\gamma
_{0}^{\prime }$,%
\begin{gather}
E=E^{\mu }\gamma _{\mu }=G[\beta _{Q}\gamma _{Q}^{2}(x^{1}-\beta
_{Q}x^{0})\gamma _{0}+\gamma _{Q}^{2}(x^{1}-\beta _{Q}x^{0})\gamma _{1}+
\notag \\
x^{2}\gamma _{2}+x^{3}\gamma _{3}],\quad B=B^{\mu }\gamma _{\mu }=0,
\label{bs}
\end{gather}%
where $G$ is that one from (\ref{fsq}). The result (\ref{bs}) significantly
differs from the result obtained by the UT, Eqs. (12a), (12b) in [1]. Under
the LT the electric field vector transforms again to the electric field
vector and the same for the magnetic field vector. It is worth mentioning
that, in contrast to the conventional results, it holds that $E^{\prime \mu
}\gamma _{\mu }^{\prime }$ \emph{from} (\ref{bf}) \emph{is} $=E^{\mu }\gamma
_{\mu }$ \emph{from} (\ref{bs}); \emph{they are the same quantity} $E$ \emph{%
for all relatively moving inertial observers.} The same holds for $B$, $%
B^{\prime \mu }\gamma _{\mu }^{\prime }$ from (\ref{bf}) is $=B^{\mu }\gamma
_{\mu }$ from (\ref{bs}) and they are $=0$ for all observers. Furthermore,
observe that in $S^{\prime }$ there are only the spatial components $%
E^{\prime i}$, whereas in $S$ there is also the temporal component $E^{0}$
as the consequence of the LT.\bigskip

\noindent \textit{5.6.2.2 The \textquotedblleft fiducial\textquotedblright\
observers are in }$S$, $v=c\gamma _{0}$, \textit{in which the charge} $Q$
\textit{is moving}$\bigskip $

Now, let us take that the \textquotedblleft fiducial\textquotedblright\
observers are in $S$, $v=c\gamma _{0}$, in which the charge $Q$ is moving, $%
u_{Q}^{\mu }=(\gamma _{Q}c,\beta _{Q}\gamma _{Q}c,0,0)$. In contrast to the
previous case, \emph{both} $E$ \emph{and} $B$ are different from zero. The
expressions for the CBGQs $E^{\mu }\gamma _{\mu }$ and $B^{\mu }\gamma _{\mu
}$ in $S$ can be simply obtained from (\ref{ecb}) taking in it that $%
v=c\gamma _{0}$ and $u_{Q}^{\mu }=\gamma _{Q}c\gamma _{0}+\beta _{Q}\gamma
_{Q}c\gamma _{1}$. This yields that $E^{0}=B^{0}=0$ (from $v=c\gamma _{0}$)
and the spatial parts are
\begin{eqnarray}
E &=&E^{i}\gamma _{i}=G\gamma _{Q}[(x^{1}-\beta _{Q}x^{0})\gamma
_{1}+x^{2}\gamma _{2}+x^{3}\gamma _{3}],  \notag \\
B &=&B^{i}\gamma _{i}=(G/c)[0\gamma _{1}-\beta _{Q}\gamma _{Q}x^{3}\gamma
_{2}+\beta _{Q}\gamma _{Q}x^{2}\gamma _{3}],  \label{seb}
\end{eqnarray}%
where $G$ is again as in (\ref{fsq}). The 4D vector fields $E$ and $B$ from (%
\ref{seb}) can be compared with the usual expressions for the 3D fields $%
\mathbf{E}$ and $\mathbf{B}$ of an uniformly moving charge, e.g., from Eqs.
(12a), (12b) in [1]. It is visible that they are similar, but $E$ and $B$ in
(\ref{seb}) are the 4D fields and all quantities in (\ref{seb}) are
correctly defined in the 4D spacetime, which transform by the LT, whereas
the fields in Eqs. (12a), (12b) in [1] are the 3D fields that transform
according to the UT.

In order to find the representations of $E$ and $B$ in $S^{\prime }$, i.e.,
the CBGQs $E^{\prime \mu }\gamma _{\mu }^{\prime }$ and $B^{\prime \mu
}\gamma _{\mu }^{\prime }$, we can either perform the LT of $E^{\mu }\gamma
_{\mu }$ and $B^{\mu }\gamma _{\mu }$ that are given by (\ref{seb}), or
simply to take in (\ref{ecb}) that relative to $S^{\prime }$ the
\textquotedblleft fiducial\textquotedblright\ observers are moving with $%
v=v^{\prime \mu }\gamma _{\mu }^{\prime }$, $v^{\prime \mu }=(c\gamma
_{Q},-\beta _{Q}\gamma _{Q}c,0,0)$, and the charge $Q$ is at rest relative
to the observers in $S^{\prime }$, $u_{Q}^{\prime \mu }=(c,0,0,0)$. This
yields the CBGQs $E^{\prime \mu }\gamma _{\mu }^{\prime }$ and $B^{\prime
\mu }\gamma _{\mu }^{\prime }$ in $S^{\prime }$ with the condition that the
\textquotedblleft fiducial\textquotedblright\ observers are in $S$, $%
v=c\gamma _{0}$,%
\begin{eqnarray}
E &=&E^{\prime \mu }\gamma _{\mu }^{\prime }=G\gamma _{Q}[-\beta
_{Q}x^{\prime 1}\gamma _{0}^{\prime }+x^{\prime 1}\gamma _{1}^{\prime
}+x^{\prime 2}\gamma _{2}^{\prime }+x^{\prime 3}\gamma _{3}^{\prime }],
\notag \\
B &=&B^{\prime \mu }\gamma _{\mu }^{\prime }=(G/c)[0\gamma _{0}^{\prime
}+0\gamma _{1}^{\prime }-\beta _{Q}\gamma _{Q}x^{\prime 3}\gamma
_{2}^{\prime }+\beta _{Q}\gamma _{Q}x^{\prime 2}\gamma _{3}^{\prime }],
\label{sc}
\end{eqnarray}%
where $G$ is again as in (\ref{bf}). Again, as in the case that $v=c\gamma
_{0}^{\prime }$, it holds that $E^{\mu }\gamma _{\mu }$ from (\ref{seb}) is $%
=E^{\prime \mu }\gamma _{\mu }^{\prime }$ from (\ref{sc}); they are the same
quantity $E$ for all relatively moving inertial observers. The same holds
for $B^{\mu }\gamma _{\mu }$ from (\ref{seb}) which is $=B^{\prime \mu
}\gamma _{\mu }^{\prime }$ from (\ref{sc}) and \emph{they are both different
from zero}. Note that in this case there are only the spatial components $%
E^{i}$ in $S$, whereas in $S^{\prime }$ there is also the temporal component
$E^{\prime 0}$ as the consequence of the LT. It is visible from (\ref{sc})
that if the $\gamma _{0}$-frame is the lab frame ($v=c\gamma _{0}$) in which
the charge $Q$ is moving then $E^{\prime \mu }\gamma _{\mu }^{\prime }$ and $%
B^{\prime \mu }\gamma _{\mu }^{\prime }$ in the rest frame of the charge $Q$%
, the $S^{\prime }$ frame, are completely different than those from (\ref{bf}%
); in (\ref{sc}) $B^{\prime \mu }\gamma _{\mu }^{\prime }$ is different from
zero and the representation of $E$\ contains also the term $E^{\prime
0}\gamma _{0}^{\prime }$.

It has to be emphasized that all four expressions for $E$ and $B$, (\ref{bf}%
), (\ref{bs}), (\ref{seb}) and (\ref{sc}), are the special cases of $E$ and $%
B$ given by (\ref{ecb}), i.e., they are different representations (CBGQs) of
$E$ and $B$ from (\ref{ebe}). They all give the same $F$ from (\ref{cf}),
which is the representation (CBGQ) of $F$ given by the basis free, abstract,
bivector (\ref{cvf}).\bigskip

\noindent \textbf{6. Conclusions\bigskip }

The whole consideration shows that in the ISR, i.e., in the approach with 4D
geometric quantities, the principle of relativity is naturally satisfied and
there is no trouble with any quantity and no paradox, i.e., that the ISR is
perfectly suited to the symmetry of the 4D spacetime, which is not the case
with the conventional SR, e.g., [10, 25 - 30], [1, 3-5], that deals with the
3D quantities and their UT or, as in the usual covariant approaches, with
components implicitly taken in the standard basis.

In the 4D spacetime, as seen from the treatment of Mansuripur's paradox that
is presented here and from the similar treatments of Jackson's paradox [11]\
and the Trouton-Noble paradox [12, 13], the \emph{physical} angular momentum
is not the 3-vector $\mathbf{L}=\mathbf{r}\times \mathbf{p}$ and the \emph{%
physical} torque is not $\mathbf{T}=\mathbf{r}\times \mathbf{F}$ with $%
\mathbf{T}=d\mathbf{L}/dt$, but \emph{the physical quantities are the 4D
geometric quantities}, $J$, i.e., $J_{s}$ and $J_{t}$ taken together, which
are given by Eqs. (\ref{jot}) - (\ref{jst}), then $N$, i.e., $N_{s}$ and $%
N_{t}$ taken together, which are given by Eqs. (\ref{en}) - (\ref{nst}). The
relation $\mathbf{T}=d\mathbf{L}/dt$ describes the usual 3D rotation, but in
the 4D spacetime it is without well-defined physical sense and it is
replaced with mathematically and relativistically correct relation (\ref{nj}%
), $N=dJ/d\tau $. According to the UT (\ref{te}) the components of the 3D
torque $\mathbf{T}$, $T_{i}$, i.e., the \textquotedblleft
space-space\textquotedblright\ components of $N^{\mu \nu }$ in one frame are
expressed by the mixture of the components of $\mathbf{T}^{\prime }$, $%
T_{i}^{\prime }$, and of another 3D quantity $\mathbf{R}^{\prime }$, $%
R_{i}^{\prime }$, i.e., the \textquotedblleft time-space\textquotedblright\
components of $N^{\prime \mu \nu }$ from a relatively moving frame.
Furthermore, if instead of Einstein's synchronization one uses a nonstandard
\textquotedblleft radio\textquotedblright\ synchronization then, even in one
frame, e.g., the \textquotedblleft time-space\textquotedblright\ components
of $N^{\mu \nu }$ in the $\{r_{\mu }\}$ basis (with the \textquotedblleft
radio\textquotedblright\ synchronization) are expressed by the mixture of
the \textquotedblleft time-space\textquotedblright\ components and the
\textquotedblleft space-space\textquotedblright\ components from the $%
\{\gamma _{\mu }\}$ basis, see similar equations for $F^{\mu \nu }$, Eqs. (%
\ref{are}) and (\ref{fe}). This proves that the 3D quantities $\mathbf{T}$
and $\mathbf{L}$ and the usual 3D rotation are not physically well-defined
in the 4D spacetime. Therefore, \emph{all treatments and all
\textquotedblleft resolutions\textquotedblright\ of Mansuripur's paradox
from [1, 3-5] are not relativistically correct in the 4D spacetime.}

Regarding the measurements of the 4D quantities, it has to be pointed out
that in the usual approaches with the 3D quantities, e.g., in the usual 3D
rotation, the experimentalists measure only three components of the 3D
torque $\mathbf{T}$ and three components of the 3D angular momentum $\mathbf{%
L}$ in both frames $S^{\prime }$ and $S$. In the 4D spacetime, the
experimentalists have to measure \emph{all six independent components} of $%
N^{\mu \nu }$ (or, equivalently, three independent components of $N_{s}^{\mu
}$ and three independent components of $N_{t}^{\mu }$), and also of $J^{\mu
\nu }$ ($J_{s}^{\mu }$ and $J_{t}^{\mu }$), in both frames $S^{\prime }$ and
$S$. The observers in relatively moving inertial frames of reference, here
in $S^{\prime }$ and $S$, are able to compare only such complete set of data
which corresponds to the \emph{same} 4D geometric quantity. It is shown in
Section 2.5 in [12] how $F$\ can be experimentally determined using the
definitions of the Lorentz force (with $F$) from Eqs. (\ref{fl}) and (\ref%
{lo}).

It is worth mentioning that different experiments for the detection of the
electric field from a stationary current loop are discussed in Section 7.2
in [2]. Recently, the most promising experiments with cold ions are proposed
in [38]. It is suggested in [2] that they could be also used for the
detection of the electric field from a stationary permanent magnet.\bigskip

\noindent \textbf{Acknowledgments\bigskip }

I am cordially thankful to Zbigniew Oziewicz for numerous and very useful
discussions during years. I am also grateful to Larry Horwitz and Alex
Gersten for the valuable discussions and for the continuos support of my
work.\bigskip

\noindent \textbf{References\bigskip }

\noindent \lbrack 1] M. Mansuripur, Phys. Rev. Lett. 108 (2012) 193901.

\noindent \lbrack 2] T. Ivezi\'{c}, J. Phys.: Conf. Ser. 437 (2013) 012014;
arXiv: 1204.5137.

\noindent \lbrack 3] A. Cho, Science 336 (2012) 404.

\noindent \lbrack 4] D. A. T. Vanzella, Phys. Rev. Lett. 110 (2013) 089401;

S. M. Barnett, Phys. Rev. Lett. 110 (2013) 089402;

P. L. Saldanha, Phys. Rev. Lett. 110 (2013) 089403;

M. Khorrami, Phys. Rev. Lett. 110 (2013) 089404;

A. Cho, http://scim.ag/Lorpara (2013); A. L. Kholmetskii, O. V. Missevitch,

T. Yarman, Prog. Electromagnetics Research B 45 (2012) 83;

T. H. Boyer, Am. J. Phys. 80 (2012) 962; C. S. Unnikrishnan, arXiv:

1205.1080; D. J. Griffiths and V. Hnizdo, arXiv: 1303.0732; arXiv: 1205.4646;

K. T. McDonald, www.physics.princeton.edu/mcdonald/examples/mansuripur.pdf;

D. J. Cross, arXiv: 1205.5451; M. Brachet and E. Tirapegui, arXiv: 1207.4613;

K. A. Milton, G. Meille, arXiv: 1208.4826;

F. De Zela, arXiv: 1210.7344; M. R. C. Mahdy, arXiv: 1211.0155.

\noindent \lbrack 5] M. Mansuripur, Phys. Rev. Lett. 110 (2013) 089405;
Proc. SPIE 8455

(2012) 845512.

\noindent \lbrack 6] T. Ivezi\'{c}, Found. Phys. Lett. 12 (1999) 507; arXiv:
physics/0102014.

\noindent \lbrack 7] T. Ivezi\'{c}, Found. Phys. 31 (2001) 1139.

\noindent \lbrack 8] T. Ivezi\'{c}, Found. Phys. Lett. 15 (2002) 27.

\noindent \lbrack 9] T. Ivezi\'{c}, arXiv: physics/0103026; arXiv:
physics/0101091.

\noindent \lbrack 10] J. D. Jackson, Classical Electrodynamics, third ed.,
Wiley, New York, 1998.

\noindent \lbrack 11] T. Ivezi\'{c}, Found. Phys. 36 (2006) 1511; Fizika A
16 (2007) 207 .

\noindent \lbrack 12] T. Ivezi\'{c}, Found. Phys. Lett. 18 (2005) 401.

\noindent \lbrack 13] T. Ivezi\'{c}, Found. Phys. 37 (2007) 747.

\noindent \lbrack 14] A. Einstein, Ann. Phys. 17 (1905) 891.

\noindent \lbrack 15] T. Ivezi\'{c}, Phys. Scr.\textit{\ }81 (2010) 025001.

\noindent \lbrack 16] J. D. Jackson, Am. J. Phys. 72 (2004) 1484.

\noindent \lbrack 17] T. Ivezi\'{c}, Found. Phys\textit{.} 33 (2003) 1339.

\noindent \lbrack 18] T. Ivezi\'{c}, Found. Phys. Lett. 18 (2005)\textbf{\ }%
301.

\noindent \lbrack 19] T. Ivezi\'{c}, Found. Phys. 35 (2005) 1585.

\noindent \lbrack 20] T. Ivezi\'{c}, arXiv: 0809.5277.

\noindent \lbrack 21] T. Ivezi\'{c}, Phys. Scr. 82 (2010) 055007.

\noindent \lbrack 22] T. Ivezi\'{c}, Phys. Rev. Lett. 98 (2007) 108901.

\noindent \lbrack 23] H. Minkowski, Nachr. Ges. Wiss. G\"{o}ttingen, 53
(1908);

Reprinted in: Math. Ann. 68 (1910) 472,

English translation in: M. N. Saha and S. N. Bose, The Principle

of Relativity: Original Papers by A. Einstein and H. Minkowski,

Calcutta University Press, Calcutta, 1920.

\noindent \lbrack 24] Z. Oziewicz, J. Phys.: Conf. Ser. 330 (2011) 012012;

Z. Oziewicz, Rev. Bull. Calcutta Math. Soc.\textit{\ }16 (2008) 49;

Z. Oziewicz and C. K. Whitney, Proc. Nat. Phil. Alliance\textit{\ }

(NPA) 5 (2008) 183 (also at http://www.worldnpa.org/php/).

\noindent \lbrack 25] D. J. Griffiths, Introduction to Electrodynamics,
fourth ed., Pearson,

Boston, 2013.

\noindent \lbrack 26] E.M. Purcell, Electricity and Magnetism,\textit{\ }2nd
ed., McGraw-Hill,

New York, 1985.

\noindent \lbrack 27] R. P. Feynman, R. B. Leighton and M. Sands, The
Feynman Lectures on

Physics Volume II, Addison-Wesley, Reading, 1964.

\noindent \lbrack 28] W. Rindler, Essential Relativity, 2nd ed.,
Springer-Verlag, New York, 1977.

\noindent \lbrack 29] W. G. W. Rosser, Classical Electromagnetism via
Relativity,

Plenum, New York, 1968.

\noindent \lbrack 30] W. K. H. Panofsky and M. Phillips, Classical
electricity and magnetism,

2nd ed., Addison-Wesley, Reading, 1962.

\noindent \lbrack 31] T. Ivezi\'{c}, arXiv: 1101.3292.

\noindent \lbrack 32] T. Ivezi\'{c}, Int. J. Mod. Phys. B 26 (2012) 1250040.

\noindent \lbrack 33] F. W. Hehl and Yu. N. Obukhov, Foundations of Classical

Electrodynamics: Charge, flux, and metric, Birkh\"{a}user, Boston, 2003.

\noindent \lbrack 34] T. Ivezi\'{c}, Found. Phys. Lett. 12 (1999) 105.

\noindent \lbrack 35] F. Rohrlich, Nuovo Cimento B 45, 76 (1966).

\noindent \lbrack 36] A. Gamba, Am. J. Phys. 35, 83 (1967).

\noindent \lbrack 37] R. M. Wald, General relativity, The University of
Chicago Press,

Chicago, 1984.

\noindent \lbrack 38] R. Folman, J. Phys.: Conf. Ser. 437 (2013) 012013;
arXiv: 1109.2586.

\end{document}